\documentclass[prc,aps,floatfix,twocolumn,showpacs,nofootinbib,superscriptaddress]{revtex4}

\usepackage{amsmath}
\usepackage{amssymb}
\usepackage{amsfonts}
\usepackage{graphicx}
\usepackage{color}

\allowdisplaybreaks     

\newcommand{\nuc}[2]{$^{#1}${#2}}

\newcommand{\nn}{\nonumber}

\newcommand{\emb}{{\rule{0cm}{0cm}}}

\begin{document}

\title{What can be learned from binding energy differences
       about nuclear structure:\\
       the example of $\delta V_{pn}$}

\author{M. Bender}
\affiliation{Universit{\'e} Bordeaux,
             Centre d'Etudes Nucl{\'e}aires de Bordeaux Gradignan, UMR5797,
             F-33175 Gradignan, France}
\affiliation{CNRS/IN2P3,
             Centre d'Etudes Nucl{\'e}aires de Bordeaux Gradignan, UMR5797,
             F-33175 Gradignan, France}

\author{P.-H. Heenen}
\affiliation{PNTPM, CP229,
             Universit{\'e} Libre de Bruxelles,
             B-1050 Bruxelles,
             Belgium}

%
%
\begin{abstract}

We perform an analysis of a binding energy difference called
$\delta V_{pn}(N,Z) \equiv - \tfrac{1}{4} \big[ E(Z,N) - E(Z,N-2)
- E(Z-2,N) + E(Z-2,N-2) \big]$ in the framework of a realistic
nuclear model.
It has been suggested that $\delta V_{pn}$ values provide a sensitive
probe of nuclear structure, and it has been put forward as a primary
motivation for the measurement of specific nuclear masses.
Using the angular-momentum and particle-number projected generator
coordinate method and the Skyrme interaction SLy4, we analyze the
contribution brought to $\delta V_{pn}$ by static deformation and dynamic
fluctuations around the mean-field ground state.
Our method gives a good overall description of $\delta V_{pn}$
throughout the chart of nuclei with the exception of the anomaly
related to the Wigner energy along the $N=Z$ line.
The main conclusions of our analysis of $\delta V_{pn}$, which are
at variance with its standard interpretation, are that
(i) the structures seen in the systematics of $\delta V_{pn}$
throughout the chart of nuclei can be easily explained combining
a smooth background related to the symmetry energy and
correlation energies due to deformation and collective fluctuations;
(ii) the characteristic pattern of $\delta V_{pn}$ having a much
larger size for nuclei that add only particles or only holes to
a doubly-magic nucleus than for nuclei that add particles for
one nucleon species and holes for the other is a trivial consequence
of the asymmetric definition of $\delta V_{pn}$, and not due to a the
different structure of these nuclei;
(iii) $\delta V_{pn}$ does not provide a very
reliable indicator for structural changes;
(iv) $\delta V_{pn}$ does not provide a reliable measure of the
proton-neutron interaction in the nuclear EDF, neither of that
between the last filled orbits, nor of the one summed over all orbits;
(v) $\delta V_{pn}$ does not provide a conclusive benchmark
for nuclear EDF methods that is superior or complementary
to other mass filters such as two-nucleon separation energies
or $Q$ values.

\end{abstract}


\date{7 February 2011}

\maketitle

%
%
\section{Introduction}

Nuclear masses are measured today with an
unprecedented accuracy \cite{Lun03a,Klu04a,Bla06a,Klu10a},
in many cases better than a few keV. Such an accuracy is
obtained not only for nuclei close to the stability line, but also
for exotic ones with very short lifetimes. A recurrent question
is how to take advantage of this major advance and how to use it to
improve the theoretical description of nuclear ground states.

The first possibility is to compare directly masses, or better binding
energies, to theoretical predictions. Unfortunately, \emph{ab-initio}
methods based on a realistic
nucleon-nucleon interaction are not available
for systematic studies of heavy nuclei. If they were, any
disagreement with the experimental data would point to a
deficiency of the interaction. When the many-body problem is
not solved exactly (or, to be more precise, with a controlled
numerical accuracy), but with an effective model using effective
degrees of freedom and an effective interaction, the link between data and
nucleon-nucleon interaction is
broken and a discrepancy between calculation and experiment can
have its source in any ingredient of the model.

The best available theoretical descriptions of masses
\cite{Lun03a,Duf,Mol,Gor,Gor09} are not based on \emph{ab-initio} methods.
The three main models rely on very different ingredients.
The mass formula of Duflo and Zuker \cite{Duf} does not
make an explicit reference to a nucleon-nucleon interaction.
Nevertheless, it assumes that there exist effective interactions
smooth enough for Hartree-Fock calculations to be possible.
The corresponding Hamiltonian is separated into  monopole
and multipole terms that are parameterized through scaling and
symmetry arguments~\cite{Lun03a}.
The macroscopic-microscopic approaches
of M{\"o}ller \emph{et al}.~\cite{Mol} combine a finite-range
liquid-drop or droplet model and shell effects introduced through
the Strutinsky shell correction method and a parameterized
one-body potential. The main ingredient of the Hartree-Fock-Bogoliubov
(HFB) mass formulae of Goriely \emph{et al}.\ is an energy
density functional (EDF) as widely used
in self-consistent mean-field calculations. In a first variant,
a Skyrme EDF is supplemented by empirical corrections for
correlations that cannot be included in a mean field~\cite{Gor}.
In a second variant, the same Gogny interaction is used to determine
the mean field and quadrupole correlation effects beyond the
mean field through a microscopic Bohr Hamiltonian~\cite{Gor09}. The
comparison of any of these models with data can hardly allow to
extract general information about the nucleon-nucleon interaction.
Comparison with results obtained using \emph{ab-initio}
methods and realistic interactions can be made through the idealized model
of infinite nuclear matter. From such calculations, one can
extract specific parameters, such as volume and symmetry energy coefficients
for instance, corresponding to a liquid-drop
formula (LDM) fitted to masses. Here also, the connection between theory
and experiment is ambiguous. The liquid-drop model is justified by
a leptodermous expansion of the energy that cannot be expected to converge
quickly even for the heaviest nuclei~\cite{Boh76a,Rei06a}.

An apparently more appealing way to proceed is to relate differences
between binding energies of neighboring nuclei to specific features
of nuclear models, in particular to effective single-particle
energies or effective two-body matrix elements \cite{Jen84a}.
Models that fail to reproduce masses with
a good accuracy are often more reliable for mass
differences. The reason of this success is that mass residuals
$M_{\text{th}} - M_{\text{expt}}$ for adjacent nuclei are not independent
in a given mass model~\cite{Hir04a,Bar05a}.

This property has been used for a long time to associate one-nucleon
separation energies with single-particle energies, or higher-order
differences with pairing gaps. In particular, two-particle separation
energies are important indicators  of shell closures.
One has to distinguish, however, between the use of mass filters
as \emph{measures} of specific model ingredients, and their use as
\emph{signatures} of structural changes. In particular, a mass filter
cannot be expected to provide both simultaneously.
One can test a model ingredient only when all nuclei
entering a mass filter have the same structure, which
becomes increasingly improbable with the number of nuclei involved.
By contrast, the indication of a structural change (such as onset of
deformation) by a mass filter often means that the fundamental
assumptions made for its direct association with a feature of a model
are violated.

In a previous paper, we discussed the difficulties
encountered when trying to relate structures seen in the systematics
of energy differences to features of the single-particle levels~\cite{Ben08a}.
The changes in the large gaps observed in data for two-neutron
or two-proton separation energies are often interpreted as due
to the evolution of shell structure with $N$ and $Z$, and
associated with the presence of strong residual tensor
interactions~\cite{Ots05a} or a weakening of the spin-orbit
interaction in neutron-rich nuclei~\cite{Eif05a,Lal98a}.
We showed that experimental data can be explained in a coherent
way within mean-field-based models as due to a combination of the
slow modification of spherical single-particle spectra and the often
rapid variation of collective correlation effects.

In this paper, we perform a similar analysis for a mass filter
that has become fashionable and has been identified
with the proton-neutron interaction in
nuclei. Many recent experimental data have been used to interpret
a difference between the (negative) binding energies of four
even-even nuclei
defined as
\begin{eqnarray}
\label{eq:dVpn:def}
\delta V_{pn} (N,Z)
& = & - \tfrac{1}{4} \big[ E(Z,N) - E(Z,N-2)
      \nonumber \\
&   & - E(Z-2,N) + E(Z-2,N-2) \big]
      \nonumber \\
&   &
\end{eqnarray}
in terms of the effective interaction between the last occupied neutron
and proton
orbits~\cite{Jan89a,Bre90a,CBC05a,Cak06a,Bre06a,Okt06a,Sto07a,Cak10a}.
It has been suggested that $\delta V_{pn}$ values provide a sensitive
probe of nuclear structure, and it has been put forward as a primary
motivation for the measurement of specific nuclear masses
\cite{Nei09aM,Nei09bM,Che09aM,Bre10aM}.
A similar quantity has been analyzed in great detail by J{\"a}necke
\emph{et al}.~\cite{Jan72a}, but for nuclei differing by one neutron
and/or one proton only, which makes its interpretation more difficult.
Indeed, the breaking of a pair in an nucleus with an odd number of
particles modifies pairing correlations deeply and makes the structure
of its wave function significantly different from that of its even
neighbors.By comparing only nuclei with an even number
of neutrons and protons, one can hope  that the assumption of a common mean field
 is better justified. Differences between two consecutive
$\delta V_{pn}$ values have also been proposed as a measure for the Wigner
energy that leads to an anomaly of binding energies around the $N=Z$
line \cite{Isa95a,Sat97a}.

In Sect.~\ref{sect:theory:analysis}, we show how, in the framework
of the Hartree-Fock (HF) method, $\delta V_{pn}$ can be related to the
interaction between the last filled neutron and proton orbits when making
the same assumptions as those used to derive Koopman's
theorem~\cite{Mar96a}. We show that the introduction of pairing
correlations and density dependencies
complicate the relation even when making oversimplifying
assumptions about the evolution of wave functions with $N$ and $Z$.
We discuss the main effects that make any direct
identification of $\delta V_{pn}$ with a proton-neutron interaction
doubtful.
In Sect.~\ref{sect:results}, we present results obtained from
calculations using a realistic microscopic model using the Skyrme
energy density functional SLy4. We demonstrate how the successive
inclusion of correlations from spherical to deformed mean-field
calculations and further to symmetry restoration and configuration mixing
permits to improve at each step the agreement with the experimental
data, while, at the same time, losing in an ever increasing manner
the simple interpretation of $\delta V_{pn}$.
Section~\ref{sect:summary} summarizes our findings.

%
%
\section{Analysis of $\delta V_{pn}$ }
\label{sect:theory:analysis}

Surprisingly, the abundant literature discussing the relevance
of $\delta V_{pn}$ contains only very few analyses of its
relation with the proton-neutron interaction in non-schematic models.
Exceptions are the shell-model study of Heyde \emph{et al}.~\cite{Hey94a},
who underline that a simple interpretation of $\delta V_{pn}$ can only
be given when there is just one dominant orbital for
protons and neutrons each,
and the nuclear DFT study by Stoitsov \emph{et al}.~\cite{Sto07a}, who,
however, focus
their analysis  on the overall excellent reproduction of data for
$\delta V_{pn}$ with their model, rather than on the
ingredients of the model that contribute to it.
Below, we review the assumptions to be made to relate
$\delta V_{pn}$ to the effective proton-neutron interaction in
finite nuclei, and discuss their validity in the context of
realistic nuclear models based on the self-consistent mean
field and taking the entire space of occupied single-particle
orbits into account. We start with
a simple two-body Hamiltonian and discuss the corresponding
energy in the context of HF (without pairing) and HFB (with
pairing). Then, we generalize the discussion to a more
realistic effective interaction that includes three-body forces
or density dependencies.

%
%

\subsection{Frozen HF with two-body interaction}
\label{subsect:frozenHF}

Let us start from a Hamiltonian consisting of a kinetic energy term
and an antisymmetrized two-body interaction:
\begin{equation}
\label{eq:H:2body}
\hat{H}
= \sum_{i,j} t_{ij} a^{\dagger}_i a_j
  + \frac{1}{4} \sum_{i,j,m,n} \bar{v}_{ijmn} \,
    a^{\dagger}_i a^{\dagger}_j a_n a_m
\, .
\end{equation}
When limiting the $N$-body wave function to a single Slater determinant,
the minimum value of the energy is obtained by solving Hartree-Fock (HF)
equations \cite{Mar96a}. In this case the energy for a nucleus consisting of
$N$ neutrons and $Z$ protons is given by
\begin{eqnarray}
\label{eq:EHF:NZ}
E^{\text{HF}}(N,Z)
& = &   \sum_{n=1}^{N} t^{N,Z}_{nn}
      + \sum_{p=1}^{Z} t^{N,Z}_{pp}
      \nonumber \\
&   &
      + \frac{1}{2} \sum_{n,n'=1}^{N} \bar{v}^{N,Z}_{n n' n n'}
      + \frac{1}{2} \sum_{p,p'=1}^{Z} \bar{v}^{N,Z}_{p p' p p'}
      \nonumber \\
&   &
      + \sum_{n=1}^{N} \sum_{p =1}^{Z} \bar{v}^{N,Z}_{n p  n p}
\, ,
\end{eqnarray}
where $t^{N,Z}_{nn}$ are the matrix
elements of the kinetic energy operator and $\bar{v}^{N,Z}_{n n' n n'}$ of the two-body
interaction calculated in the single-particle basis that solves the
HF equations. We have added superscripts $N,Z$ to these matrix elements
to recall that the HF equations are solved self-consistently and that,
in general, the wave functions differ for each combination of $N$ and
$Z$ values.

Let us assume for the moment that the HF single-particle basis is
identical for the four nuclei with $(N,Z)$, $(N-2,Z)$, $(N,Z-2)$,
and $(N-2,Z-2)$. This ``frozen HF'' approximation leads to
\begin{eqnarray}
\lefteqn{\delta V^{\text{HF}_{\text{frozen}}}_{pn} (N,Z)
} \nonumber \\
& = & -\tfrac{1}{4} \;
     \big(   \bar{v}_{N-1, Z-1, N-1, Z-1}
           + \bar{v}_{N-1, Z  , N-1, Z  }
     \nonumber \\
&  & \quad
           + \bar{v}_{N  , Z-1, N  , Z-1}
           + \bar{v}_{N  , Z  , N  , Z  }
     \big)
\, .
\end{eqnarray}
The superscripts have been dropped, as the mean field is supposed
to be the same for the four nuclei.
\footnote{This assumption, however, cannot be expected to be valid for
all nuclei. Each nucleus enters the expression for $\delta V_{pn}$
for four different nuclei, such that ultimately \emph{all} nuclei
had to have the same mean field for this assumption to be fulfilled.}

A further simplification can be obtained by noting that, for even $N$ and
$Z$, the two valence neutrons (indices $N-1$ and $N$) and protons
(indices $Z-1$ and $Z$) occupy time-reversed orbits in the HF solution,
and that the matrix elements are equal two by two
\begin{equation}
\label{eq:dVpn:HF:frozen}
\delta V^{\text{HF}_{\text{frozen}}}_{pn} (N,Z)
 = - \frac{1}{2}
      \big(   \bar{v}_{N  , Z-1, N  , Z-1}
            + \bar{v}_{N  , Z  , N  , Z  }
      \big)
\, .
\end{equation}
Even in this simple case, the final result is not a single matrix
element, but a combination of matrix elements between the valence
particles.

This derivation requires the same assumptions as those made to derive
Koopman's theorem~\cite{Mar96a} which relates single-particle energies and
one-nucleon separation energies. Koopman's theorem, however, is known to
have a very limited validity
in nuclear physics,
{cf.\ the discussion in Ref.~\cite{Ben08a} and references therein.

%
%

\subsection{Frozen HF+BCS and HFB with two-body interaction}

It is obvious that the assumptions made in the previous section
will rarely be justified, even approximately. Let us first examine the
consequence the partial occupation of single-particle levels
due to pairing correlations. We consider only pure
proton-proton and neutron-neutron pairing, which can be justified for
nuclei with $N$ sufficiently different from $Z$. In the presence of
proton-neutron pairing, the many-body state could not be written
as the direct product of a proton and a neutron BCS state, and the
energy would not be separable into proton and neutron components
anymore.

The HF+BCS or HFB expectation value of a two-body Hamiltonian
(\ref{eq:H:2body}) for a nucleus with $N$ neutrons and $Z$ protons
evaluated in the canonical basis is given by
\begin{eqnarray}
\label{eq:EHFB:NZ}
E^{\text{HFB}}(N,Z)
& = &   \sum_{n} t^{N,Z}_{nn} \, v^2_{n,N}
      + \sum_{p} t^{N,Z}_{pp} \, v^2_{p,Z}
      \nonumber \\
&   &
      + \frac{1}{2} \sum_{n,n'}
        \bar{v}^{N,Z}_{n n' n n'} \, v^2_{n,N} \, v^2_{n',N}
      \nonumber \\
&   &
      + \sum_{n} \sum_{p}
        \bar{v}^{N,Z}_{n p  n p} \, v^2_{n,N} \, v^2_{p,Z}
      \nonumber \\
&   &
      + \frac{1}{2} \sum_{p,p'}
        \bar{v}^{N,Z}_{p p' p p'} \, v^2_{p,Z} \, v^2_{p',Z}
      \nonumber \\
&   &
      + \sum_{n,n' > 0} \bar{v}^{N,Z}_{n \bar{n} n' \bar{n}'} \,
        u_{n,N} \, v_{n,N} \, v_{n',N} \, v_{n',N}
      \nonumber \\
&   &
      + \sum_{p,p' > 0} \bar{v}^{N,Z}_{p \bar{p} p' \bar{p}'} \,
        u_{p,Z} \, v_{p,Z} \, v_{p',Z} \, v_{p',Z}
\, .
\end{eqnarray}
The $u_k$ and $v_k$ are real occupation amplitudes with
$u_k = u_{\bar{k}} > 0$, $v_k = -v_{\bar{k}}$ and $u_k^2 + v_k^2 = 1$.
Indices $k$ and $\bar{k}$ refer to conjugate states, which for
the ground states of even-even nuclei are connected by the time-reversal
operator.
We use the usual convention where summation over all indices indicates
summation over all $k$ and $\bar{k}$, whereas summation over ``positive''
indices means that the sums are over all $k$, but not their conjugate
levels $\bar{k}$.
The second index of the occupation amplitudes recalls that these are
occupation numbers for a nucleus with $\sum_{n} v^2_{n,N} = N$
neutrons and $\sum_{p} v^2_{p,Z} = Z$ protons. Except for the last
two terms, the summation runs over positive and negative values of
$n$ and $p$.

To derive a simple expression for $\delta V_{pn}$,
a similar assumption as in the frozen HF case has to be made,
namely that the canonical single-particle basis is the same for all
four nuclei entering a given $\delta V_{pn} (N,Z)$. In this case, the
matrix elements $t_{kk}$ and $\bar{v}_{kk' kk'}$ do not depend on $N$
and $Z$. In addition, one has to assume that the solution of the HFB
equations for neutron states does not depend on the
number of protons and conversely. This leads to
\begin{equation}
\label{eq:dVpn:HFB:frozen}
\delta V^{\text{HFB}_{\text{frozen}}}_{pn} (N,Z)
 =  - \frac{1}{4} \sum_{n} \sum_{p}
      \bar{v}_{n p n p} \, \Delta v^2_{n,N} \, \Delta v^2_{p,Z}
\, ,
\end{equation}
where we introduced the shorthand
$\Delta v^2_{n,N} \equiv v^2_{n,N} - v^2_{n,N-2}$ for the change of
neutron occupation numbers when removing two neutrons,
$\sum_n \Delta v^2_{n,N} = 2$, and its homologue
$\Delta v^2_{p,Z} \equiv v^2_{p,Z} - v^2_{p,Z-2}$ for protons.

%
%
\subsection{Frozen HF with three-body or density-dependent interactions }

The  forms derived in the previous sections were based on a two-body
Hamiltonian. However, interactions derived from first principles contain
at least three-body forces~\cite{Epe09a}, which are crucial to perform
\emph{ab-initio} calculations with some predictive power. In a similar
manner, the more phenomenological EDF-based methods must include terms
of higher order than
a two-body interaction, which is usually done by the inclusion
in the EDF of terms with density dependencies. In particular, the
saturation of nuclear matter cannot be satisfactorily
described without taking into account these terms~\cite{Ben03a}.

A three-body force adds a term $\tfrac{1}{36} \sum_{i,j,k,l,m,n} v_{ijklmn}
a^{\dagger}_i a^{\dagger}_j  a^{\dagger}_k a_n a_m a_l$ to the
Hamiltonian~(\ref{eq:H:2body}). In the HF approximation, its contribution
to the binding energy is given by:
\begin{eqnarray}
\label{eq:EHF:3body}
E^{\text{HF}}_{3b}(N,Z)
& = &   \frac{1}{6} \sum_{n,n',n''=1}^{N}
    \bar{v}^{N,Z}_{n n' n'' n n' n''}
      \nonumber \\
&   &
  + \frac{1}{2} \sum_{n,n'=1}^{N} \sum_{p =1}^{Z}
    \bar{v}^{N,Z}_{n n' p   n n' p  }
      \nonumber \\
&   &
  + \frac{1}{2} \sum_{n=1}^{N} \sum_{p,p' =1}^{Z}
    \bar{v}^{N,Z}_{n p  p'  n p  '' }
      \nonumber \\
&   &
  + \frac{1}{6} \sum_{p,p',p''=1}^{Z}
    \bar{v}^{N,Z}_{p p' p'' p p' p''}
\, ,
\end{eqnarray}
using a notation analogous to that of Eq.~(\ref{eq:EHF:NZ}).
In the frozen HF approximation, a Hamiltonian including two-body and
three-body forces leads to
\begin{eqnarray}
\label{eq:dVpn:fHF:2+3}
\lefteqn{
\delta V^{\text{HF}_{\text{frozen}}}_{pn} (N,Z)
} \nn \\
& = & - \frac{1}{4}
      \Bigg[
      \sum_{n=N-1}^{N} \sum_{p=Z-1}^{Z} \bar{v}_{npnp}
      \nonumber \\
&   &
      + \frac{1}{2}
        \sum_{n,n'=N-1}^{N}  \sum_{p=Z-1}^{Z}
        \bar{v}_{n n' p n n' p}
      \nonumber \\
&   &
      + \frac{1}{2}
        \sum_{n=N-1}^{N} \sum_{p,p'=Z-1}^{Z}
        \bar{v}_{n p p' n p p'}
      \nonumber \\
&   &
      + \sum_{n=1}^{N-2} \sum_{n'=N-1}^{N} \sum_{p=Z-1}^{Z}
        \bar{v}_{n n' p n n' p  }
      \nonumber \\
&   &
      + \sum_{p=1}^{Z-2} \sum_{n=N-1}^{N}  \sum_{p'=Z}^{Z-2}
        \bar{v}_{n p p' n ' p'}
      \Bigg]
\, .
\end{eqnarray}
The first three lines represent the two-body and three-body interactions
between the last two neutrons and the last two protons. The
last two lines of Eq.~(\ref{eq:dVpn:fHF:2+3}), however, contain a
sum over all other nucleons, which is incompatible with the interpretation
of $\delta V_{pn} (N,Z)$ as the interaction between the last two
neutrons and protons.

Most energy density functionals constructed for self-consistent
mean-field calculations include a non-linear density-dependent
two-body term of the form
$\tfrac{1}{4} \sum_{i,j,m,n} v_{ijkl} \, f( \rho_n + \rho_p ) \,
a^{\dagger}_i a^{\dagger}_j a_n a_m$, where $\rho_n$ and $\rho_p$
are the local densities of neutrons and protons. Popular examples
for $f ( \rho_n + \rho_p )$ range from the simple non-integer powers
of the total density $f(x) = x^\alpha$, such as used with most Skyrme
and Gogny interactions, to the very elaborate density dependencies
$f(x) \sim [1+b(x+d)^2] / [1+c(x+d)^2]$ used in modern density-dependent
relativistic mean-field models~\cite{Typ99a,Nik02a}.
Any density dependence that is not a simple polynomial of the total
density gives rise to $\delta V_{pn}$ values that cannot be separated
into neutron and proton contributions, even within the frozen HF
approximation.

We will not give explicit expressions for the HFB case with three-body
forces or density-dependent terms. It should be obvious by now
that even making the assumption of a frozen common canonical basis will
lead to a lengthy and complicated expression for $\delta V_{pn} (N,Z)$ that
does not allow for an intuitive interpretation.

%
%

\subsection{Discussion}

Even when making the drastic approximation that the four nuclei
entering the calculation of $\delta V_{pn} (N,Z)$ can be described
by the same mean field,
the expression that is obtained for realistic models
includes a summation over all single-particle levels. \footnote{
A similar observation has been made by Van Isacker \emph{et al.}~\cite{Isa95a}
who quote that "$\delta V_{pn} (N,Z)$ is an average neutron-proton interaction over the
last few nucleons", but without giving any reference.
}
Furthermore,
the occupation of the levels around the Fermi energy
is affected by the addition or the removal of nucleons and
the contribution of each single-particle level
to $\delta V^{\text{HFB}_{\text{frozen}}}_{pn}(N,Z)$
is weighted with the difference of its occupation between the nuclei.
It is therefore doubtful that $\delta V_{pn}$ allows to isolate
an empirical interaction between the last neutron and proton orbitals,
as claimed in Ref.~\cite{Bre06a}.

Moreover, one might wonder whether the approximation of a frozen canonical
basis necessary to derive Eqns.~(\ref{eq:dVpn:HF:frozen})
and~(\ref{eq:dVpn:HFB:frozen})
is ever satisfied. To change the number of neutrons or protons by two
induces rearrangement and polarization effects that modify the
single-particle wave functions for both kinds of nucleons. Even if these
effects are most often small, it should not be forgotten that
$\delta V_{pn} (N,Z)$ is a tiny
fraction of the total
binding energy only, ranging from $10^{-2}$ in light nuclei to $10^{-5}$
in heavy ones. Therefore, even small rearrangement effects
can have a large impact on the values obtained for $\delta V_{pn} (N,Z)$.
We will analyze the validity of the frozen basis assumption for a few
selected cases in Sect.~\ref{subsect:results:selectednuclei} below.
Finally, a self-consistent
mean-field description of a nucleus provides a reasonable
first approximation but it neglects correlations beyond the mean field
that also contribute to the binding
energy on the MeV scale~\cite{Ben05a,Ben06a}. These correlations cannot be cast in a
simple form involving only the interaction between a few particles,
and they also destroy the simple relation between $\delta V_{pn}$ and
proton-neutron matrix elements.

%
%

In the remaining part of this article, we investigate the
importance of self-consistency, deformation, pairing, and configuration
mixing for the description of data for $\delta V_{pn}$
We  also analyze to which extend $\delta V_{pn}$ values
can be identified with the effective proton-neutron interaction.

%
%

\section{The models}
\label{sect:models}

%
%
\subsection{The beyond-mean-field model}

Our method used to calculate binding energies for the ground states of
even-even nuclei is described in detail in Refs.~\cite{Ben05a,Ben06a}.
In our analysis, we use the energies as tabulated in~\cite{EPAPS},
and we added a few nuclei in the vicinity of \nuc{208}{Pb}.
As effective interaction, we employ the SLy4 parameterization of
the Skyrme energy density functional~\cite{Cha98a} for the
mean-field channel in connection with a density-dependent zero-range
pairing interaction.

Starting from a set of mean-field calculations including a constraint
on the axial quadrupole moment, two kinds of correlations beyond the
mean-field are introduced. First, the deformed wave functions are
projected on both fixed particle numbers and on angular momentum $J=0$.
In a collective model terminology, these correlations would be called
rotational correlations. A second step of our method consists in the
mixing of projected wave functions with different intrinsic axial
quadrupole moment of the underlying mean-field state in a generator
coordinate method (GCM).
In the language
of collective models, this corresponds to a vibrational correction.
The final wave function has the form
\begin{equation}
| J M \nu \rangle
= \sum_{q} f_{J,\nu} (q) \hat{P}^J_{M0} \hat{P}_N \hat{P}_Z | q \rangle
\, .
\end{equation}
The ket $| q \rangle$ is a (paired) self-consistent mean-field state
of axial quadrupole deformation $q$. The operators $\hat{P}_N$,
$\hat{P}_Z$ and $\hat{P}^J_{M0}$ project out the component
with the particle numbers and angular momentum quantum number
we are interested in. The weights $f_{J,\nu} (q)$ defining the
mixing of the projected wave functions with respect to $q$ are
obtained by variation of the total energy.

We stress that there are no assumptions made in the model about the
amplitude of the quadrupole fluctuations introduced into the calculations.
Depending on the structure of a nucleus, this amplitude either corresponds
to a small vibration around a pronounced minimum, to a large-amplitude
motion in a soft and wide potential well, or to the mixing of several
states around coexisting minima in the deformation energy surface.

In the following, we will compare results obtained from the energies
determined using three wave functions that successively add quadrupole
correlations:
\begin{enumerate}
\item
self-consistent spherical mean-field states $| q=0 \rangle$;
\item
the self-consistent mean-field minimum in the space of axial
reflection-symmetric deformations  $| q_{\text{min}} \rangle$,
which might be spherical;
\item
the ground state obtained after configuration mixing of $J=0$ projected
axial quadrupole. We refer to these wave functions in the following as
projected GCM. The energy gained through these correlations will be
called beyond-mean-field correlation energy in what follows.
\end{enumerate}
In each of these cases, the wave functions are projected on particle number.

\begin{figure}[t!]
\centerline{\includegraphics[height=19.8cm]{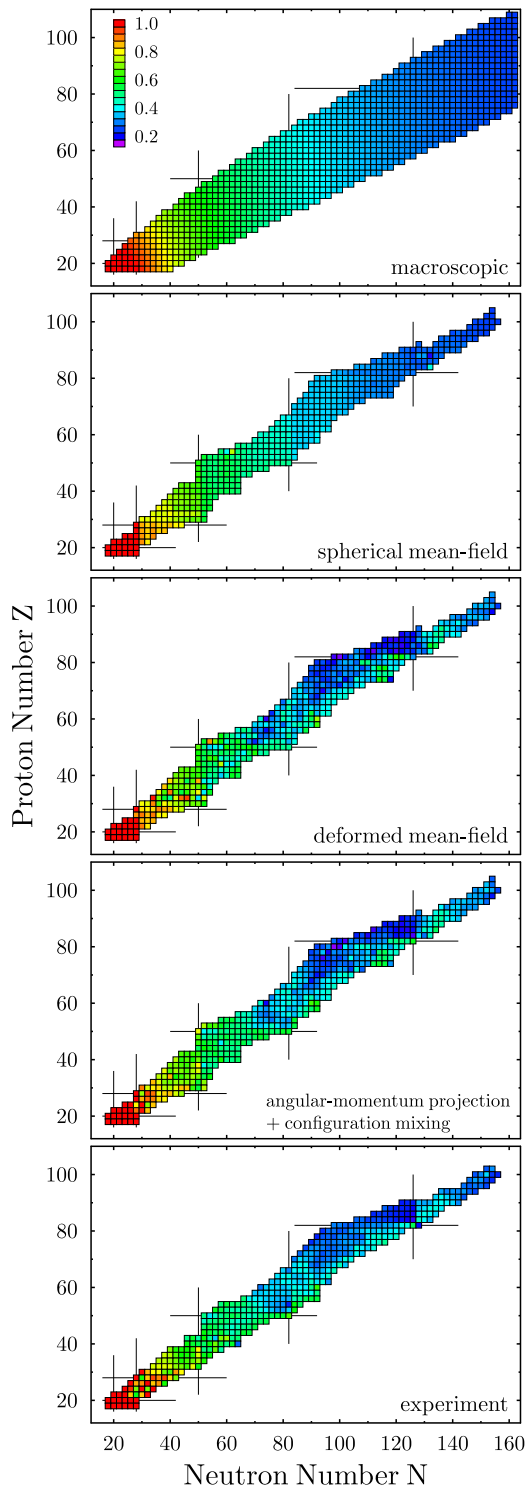}}
\caption{
\label{fig:dvpn_map}
(Color online)
Maps of $\delta V_{pn}$; From top to bottom: calculated
with a spherical liquid drop formula having the average properties of the
SLy4 interaction, by self-consistent calculations with SLy4 obtained assuming spherical
nuclei, allowing for (axially) deformed shapes, and derived from $J=0$ projected
configuration mixing calculations. The bottom panel shows experimental
data.
}
\end{figure}

%
%
\subsection{Liquid drop model}

The Strutinsky theorem \cite{Str67a,Bra75a} allows to decompose
the binding energy into a ``macroscopic'' liquid drop part and a
microscopic ``shell correction''. In this picture,
the macroscopic energy defined through the liquid drop model
varies smoothly with $N$ and $Z$, without
any correlation energies from deformation, shell effects, or
fluctuations in collective degrees of freedom. It constitutes a reference
with respect to which one can put into evidence all quantum effects.

In some of the figures below we show macroscopic energies
calculated from a liquid-drop model whose parameters have been adjusted
to reproduce the average binding energies of spherical nuclei
calculated with the Skyrme interaction SLy4 \cite{Rei06a}.
Besides the standard volume, volume symmetry, surface and (direct) Coulomb
terms, the macroscopic model comprises surface symmetry, curvature and
Coulomb exchange terms
\begin{eqnarray}
\label{eq:mac}
E_{\text{mac}} (N,Z)
& = &   ( a_{\text{vol}}  + a_{\text{sym}} \, I^2 ) \, A
      \nonumber \\
&   &
      + ( a_{\text{surf}} + a_{\text{surf,sym}} \, I^2 ) A^{2/3}
      \nonumber \\
&   &
      +   a_{\text{curf}}  A^{1/3}
      \nonumber \\
&   &
      + \frac{3}{5} \frac{Z^2 e^2}{r_0 A^{1/3}}
      - \frac{3}{4} \left( \frac{3}{2\pi}\right)^{2/3}
        \frac{Z^{4/3} e^2}{r_0 A^{1/3}}
      \nonumber \\
&   &
\end{eqnarray}
where $A = N+Z$ and $I=\frac{N-Z}{N+Z}$. The radius constant $r_0$ entering
the Coulomb energies is determined from the nuclear matter saturation density
$\rho_0$ of SLy4 as $r_0^3 = 3/(4 \pi \rho_0)$.

The by far dominating contribution to $\delta V_{pn}$ comes from the
volume and surface symmetry energies \cite{Sto07a}
\begin{equation}
\label{eq:mac:sym}
\delta V_{pn}
\approx 2 \big( a_{\text{sym}} + a_{\text{surf,sym}} A^{-1/3} \big) \, A^{-1}
\, .
\end{equation}
The global $A^{-1}$ scaling factor in this expression originates from the
denominator of the $I^2 \, A = \frac{(N-Z)^2}{N+Z}$ factor in the symmetry
and surface symmetry energy terms which do not cancel out in
$\delta V_{pn}$.
%
%

There are two contributions to this term which have the same
scaling~\cite{Heyde}: the first one is the difference in kinetic energy
between protons and neutrons that fill separate potential wells, and the
other the isovector part of the nucleon-nucleon interaction. The latter
has a shorter range than the average distance between nucleons, such that
in a semiclassical approximation it acts between nearest neighbors only, leading
to the characteristic $A$ and $A^{1/3}$ scaling of terms in
Eq.~(\ref{eq:mac:sym}).
The contribution of all other terms in Eq.~(\ref{eq:mac}) to $\delta V_{pn}$
is not exactly zero, but it is too small to be resolved in the plots shown
below. A standard liquid-drop model obtained with a ``best fit'' to
experimental masses gives $\delta V_{pn}$ values that are systematically
larger than those obtained from Eq.~(\ref{eq:mac}), mainly because the
volume symmetry coefficient $a_{\text{sym}}$ has a slightly larger value
than the one determined from the SLy4 interaction.

%
%
\section{Results}
\label{sect:results}

%
%
\subsection{Global behavior of $\delta V_{pn}$}
\label{subsect:results:global}

%
%

The  binding energies of Refs.~\cite{Ben05a,Ben06a}
and tabulated in~\cite{EPAPS} cover the region of even-even nuclei heavier
than \nuc{16}{O} for which experimental data are available, plus a few
additional nuclei around doubly-magic systems. For the present study, we
calculated a few extra nuclei around \nuc{208}{Pb}. Values obtained
for $\delta V_{pn}$ with this sample of nuclei are plotted as maps
in Fig.~\ref{fig:dvpn_map}.
For a better resolution of the local fluctuations, the same data are
plotted for isotopic chains as a function of the number of neutrons
in Fig.~\ref{fig:dvpn_n}, and for isotonic chains as a function
of proton number in Fig.~\ref{fig:dvpn_p}.

The spherical macroscopic values are given each time in the top panel.
All nuclei between the drip lines are represented for the LDM results
in Fig.~\ref{fig:dvpn_map}, whereas
in Figs.~\ref{fig:dvpn_n} and~\ref{fig:dvpn_p} results are restricted
to the same set of nuclei shown in the other panels.
The macroscopic $\delta V_{pn}$ values exhibit a regular smooth
pattern and fall off with $\sim 1/A$. The slope of
the decrease is related to the symmetry and surface symmetry energy
coefficients of the EDF, Eq.~(\ref{eq:mac:sym}).
This smooth systematic decrease of the $\delta V_{pn}$ values with
increasing $A$ has been sometimes interpreted
as a result of ``the gradual decrease in valence proton and neutron
orbital overlaps due to the occupancy of shells of different average
radii'' \cite{Bre90a,Bre06a}. This is, at best, a model-dependent
statement that cannot be translated to methods that calculate the
energy from the interaction between all occupied particles.
In particular, at no point in the derivation of the LDM
expression~(\ref{eq:mac:sym})
has one to consider the form of the single-particle wave functions
and their overlaps. Instead, it is only assumed that all
occupied single-particle wave functions add up to the saturation
 density inside the nucleus.

\begin{figure}[t!]
\centerline{\includegraphics{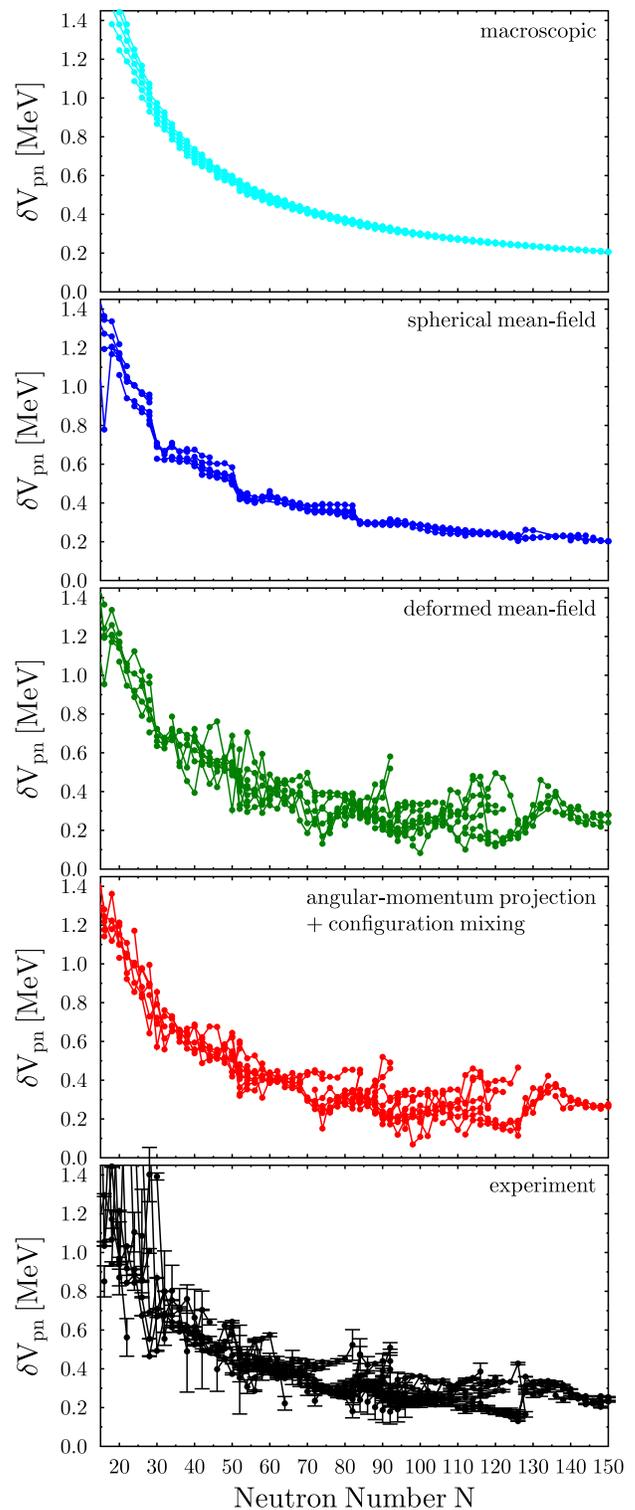}}
\caption{
\label{fig:dvpn_n}
(Color online)
Same data as Fig.~\ref{fig:dvpn_map}, but plotted for isotopic
chains as a function of neutron number.
}
\end{figure}

\begin{figure}[t!]
\centerline{\includegraphics{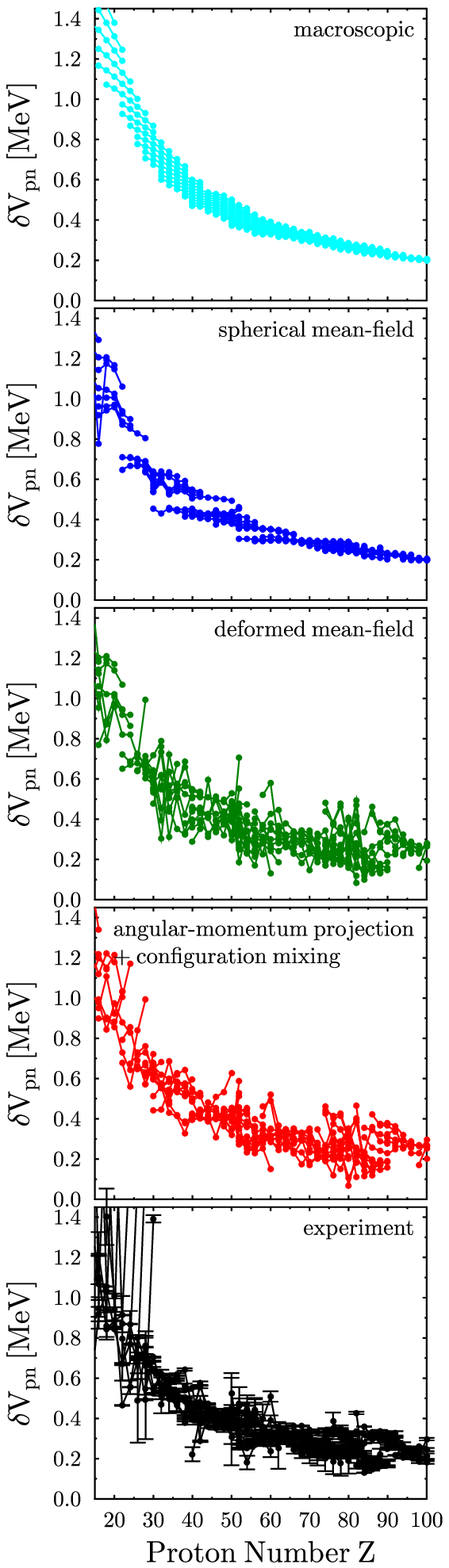}}
\caption{
\label{fig:dvpn_p}
(Color online)
Same data as Fig.~\ref{fig:dvpn_map}, but plotted for isotonic
chains as a function of proton number.
}
\end{figure}

\begin{figure}[t!]
\centerline{\includegraphics[width=6.7cm]{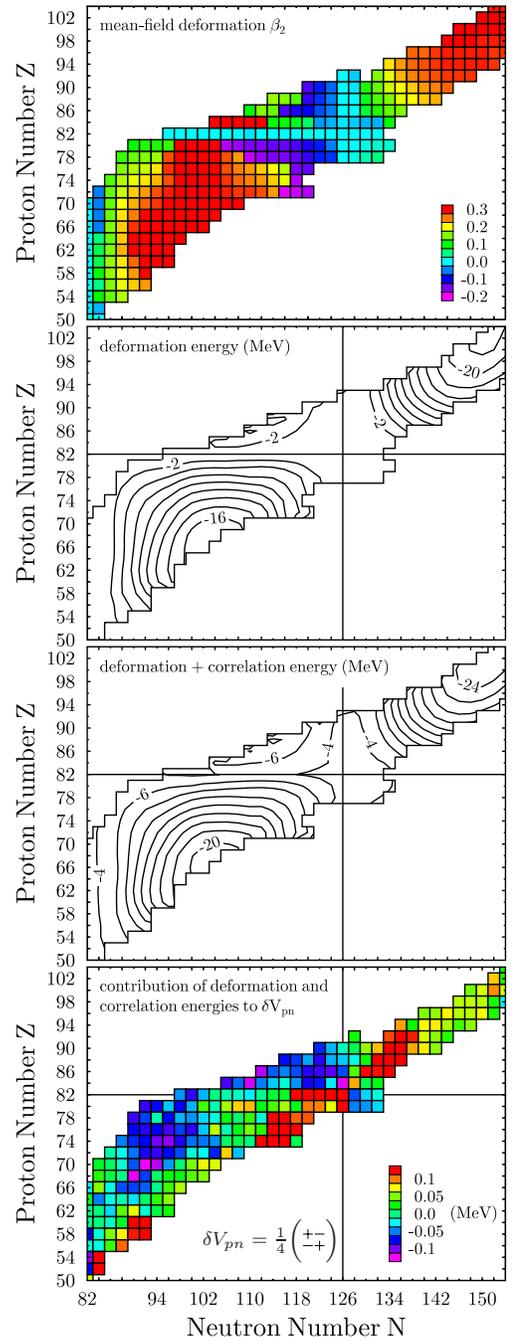}}
\caption{
\label{fig:dvpn_def}
Effect of deformation and correlation energies on $\delta V_{pn}$
values for heavy nuclei.
Top to bottom: (i) map of the dimensionless quadrupole deformation
$\beta_2$ of the mean-field ground state,
(ii) contour plot of the static deformation energy,
(iii) contour plot of the sum of the static deformation and dynamical
beyond-mean-field correlation energies, and
(iv) map of the contribution of deformation and dynamical correlation
energy to $\delta V_{pn}$. The total $\delta V_{pn}$ value is then
obtained adding the spherical mean-field value.
The inset gives a reminder of the relative signs of the
four contributions to $\delta V_{pn}$, their distance being
drawn on the same scale as the one used in the contour plot.
}
\end{figure}

The smooth trend of the macroscopic calculation is still apparent in
the spherical self-consistent mean-field results.
Some deviations appear, however, which are related to
the magic numbers at 20, 28, 50, 82 and 126. For nuclides just
below these shell closures, the spherical mean-field results
are slightly larger than the LDM ones, whereas they take
slightly smaller values for nuclei just above. As a consequence,
$\delta V_{pn}$ values do
not fall off continuously with $A$,
but form sheets separated by the shell closures.

Relaxing the constraint of spherical symmetry strongly modifies the
behavior of $\delta V_{pn}$ by giving rise to rapid fluctuations
around the
smooth trend, with an amplitude of up to 200 keV. This
change can be directly related to the effect of deformations on binding
energies. The variation of quadrupole deformation
and
of the associated energy gain with $N$ and $Z$ over the entire nuclear
chart have been presented in Figs.~9 and~16 of Ref.~\cite{Ben06a}.
The energy gain due to deformation can reach more than 20~MeV and can vary
rapidly from one nucleus to the other.
Any mismatch in the evolution of deformation energy between the four
nuclei entering $\delta V_{pn}$ can dramatically change its value.

To illustrate the impact of deformation and correlations on $\delta V_{pn}$,
the quadrupole deformation $\beta_2$, deformation and correlation energies
for nuclei heavier than \nuc{132}{Sn} are presented in Fig.~\ref{fig:dvpn_def}.
As expected, the absolute value of the
deformation energy increases first slowly when moving away
from the proton and neutron shell closures
and then more rapidly until it peaks at almost 18 MeV in the
rare-earth region and above 20 MeV for actinides.
As indicated in the inset of the figure, $\delta V_{pn}$ is defined
as the sum of the energies of the nuclei on the diagonal minus
the sum of the energies of the nuclei on the anti-diagonal.
Figure~\ref{fig:dvpn_def} gives an intuitive illustration of
how the deformation and beyond-mean-field correlation energies
contribute
to $\delta V_{pn}$. Along a line going from \nuc{132}{Sn}
to \nuc{208}{Pb} and beyond,
the deformation energy varies rapidly and nonlinearly
and brings a very large contribution to $\delta V_{pn}(N,Z)$. On the contrary,
for nuclei located close to a spherical shell closure for one
nucleon species and mid-shell for the other,
the lines of equal deformation and correlation}\emb\
energy are nearly parallel to the $N$ {or $Z$
axis, leading to a pairwise cancelation of similar deformation energies.
The fine structure of this cancelation depends of course on the deformed
shell structure of the four nuclei entering a given $\delta V_{pn}(N,Z)$ value
and leads to an erratic behavior of $\delta V_{pn}$,
particularly visible when drawn for isotopic or isotonic
chains as in Figs.~\ref{fig:dvpn_n} and~\ref{fig:dvpn_p}. The same
behavior is also seen in the experimental data.
It is worthwhile to mention that there is no direct relation between the
size of the deformation and the deformation energy, nor between the
sign of the deformation, prolate or oblate, and the deformation energy.
In particular, the transition between oblate and prolate shapes in
a region of shape coexistence around the neutron deficient Pb isotopes
does not leave obvious traces in the ground-state
deformation energy, hence, in the calculated $\delta V_{pn}$ values.
As one may expect, introducing beyond-mean-field correlations
evens out the effect of static
deformation. The effect of configuration mixing
is, indeed, a spreading of the ground-state wave function around
the mean-field minimum and a mixing of coexisting shapes.
The beyond-mean-field correlation energy varies rapidly
only around shell closures and has its largest impact
on the $\delta V_{pn}(N,Z)$ values in these regions.
We will analyze its impact in more detail for selected nuclei below.

The very rapidly varying behavior of $\delta V_{pn}$ around
the $N=Z$ line that sticks out in the experimental data for light nuclei
in Figs.~\ref{fig:dvpn_map},~\ref{fig:dvpn_n} and~\ref{fig:dvpn_p}
is not reproduced by any of our calculations. This anomaly
is due to the Wigner energy~\cite{Isa95a,Sat97a,Zel98a,Kir08a},
whose origin is not described by present-day
EDF models, see Refs.~\cite{Sat97a,Per04a} for further
discussion of this deficiency.

In the literature one cannot find, however, a unique definition of the
Wigner energy.
Sometimes this notion is used for an anomalous additional contribution to
the binding energy of the $T=0$ $\Leftrightarrow$ $N = Z$ member of an
isobaric multiplet compared to the $(N-Z)^2$ extrapolation from the other
isobars \cite{Pov98a}, but more often the Wigner energy denotes a contribution
to the binding energy that is linear in $|T_z| = |N-Z|$. Such a term arises,
for example, from a Hamiltonian that is invariant under Wigner's $SU(4)$
symmetry, $E \sim \mbox{$T(T+4)$} = E_{\text{sym}} + E_{\text{Wigner}}$.
When using this second concept of a Wigner energy, the anomaly of
binding energies at $N=Z$ is the consequence of the Wigner energy
$E_{\text{Wigner}}$ having a discontinuity in its derivative at
$T_z = 0$, and not of an additional binding of $N=Z$ nuclei. There are
many reasons why Wigner's $SU(4)$ symmetry is not realized in
nuclei~\cite{War06a,Isa95a,Sat97a,Zel98a,Kir08a}, which suppresses
the linear term in $|N-Z|$ compared to the quadratic one. Still,
traces of such a linear term are implicitly contained in all realistic
shell-model calculations~\cite{Sat97a,Cau05a,Zel98a,Pov98a} and explicitly
in the Duflo-Zuker mass formula~\cite{Duf}. It is noteworthy that the
macroscopic-microscopic mass models~\cite{Mol} and the Skyrme-HFB mass
formulae~\cite{Gor} also contain explicit phenomenological corrections for
the Wigner energy that, in fact, combine both of the concepts of Wigner
energy mentioned above, see also the discussion in Ref.~\cite{Lun03a}.
Several mass differences have been put forward as indicators or even
measures of the Wigner energy, most prominently double-$\beta$-decay
$Q$~values~\cite{Zel98a,Kir08a} or a difference between three different
$\delta V_{pn}$ values~\cite{Sat97a}. It has to be stressed, however,
that $\delta V_{pn}$ itself is not a measure of the Wigner energy.

%
%
\subsection{Selected chains of nuclei}
\label{subsect:results:chains}

%
%
\subsubsection{Isotopic chains of magic nuclei}

\begin{figure}[t!]
\centerline{\includegraphics{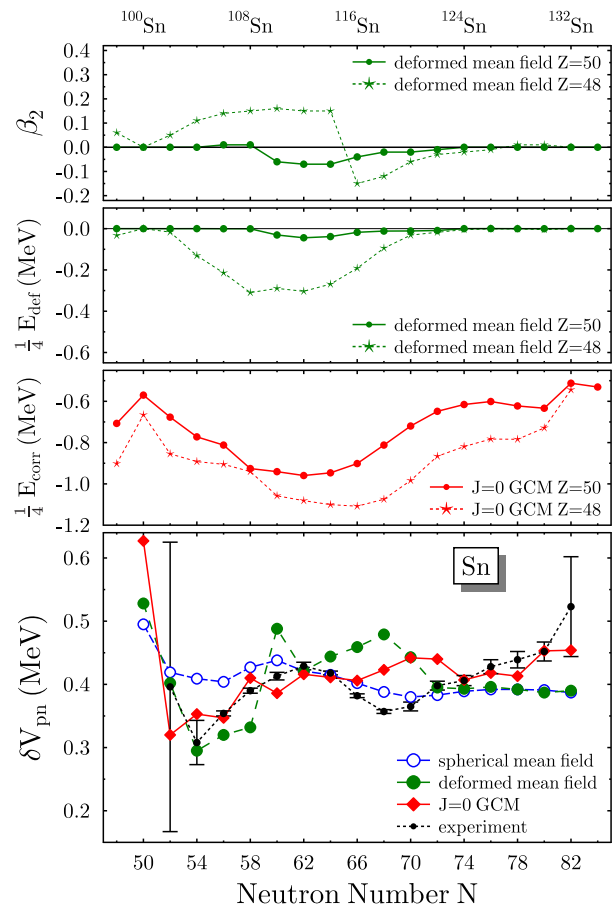}}
\caption{
\label{fig:dvpn_50}
(Color online)
Top to bottom: Intrinsic deformation of the mean-field
ground state,
$1/4$ of the deformation energy, $1/4$ of the beyond-mean-field
correlation energy, and $\delta V_{pn}$ values obtained from spherical
mean-field calculations, deformed mean-field calculations, $J=0$ projected
GCM calculations and experimental data for the chain of Sn ($Z=50$)
isotopes. For the deformation, deformation energy and correlation energy
the values for Cd ($Z=48$) isotopes are also shown.
}
\end{figure}

To demonstrate how $\delta V_{pn}$ is build up from different types of
correlations, let us examine now its evolution along cuts through the
$N$--$Z$ plane. We first look at the two isotopic chains of Sn and Pb,
corresponding to closed proton shells, in Figs.~\ref{fig:dvpn_50}
and~\ref{fig:dvpn_82}.
We showed in Ref.~\cite{Ben08a} for the $Z=50$ Sn isotopic chain
that static deformation and dynamical correlation energies are
key ingredients in reproducing the two-proton separation
energies across the $Z=50$ shell and in explaining in an intuitive
way the mutually enhanced magicity around \nuc{132}{Sn}.

\begin{figure}[t!]
\centerline{\includegraphics{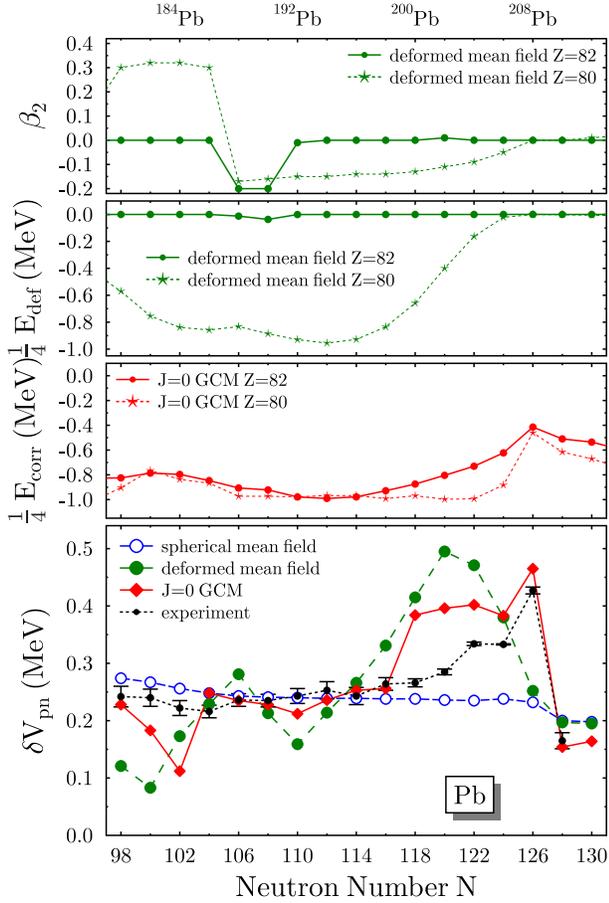}}
\caption{
\label{fig:dvpn_82}
(Color online)
The same as Fig.~\ref{fig:dvpn_50}, but for the chain of Pb
($Z=82$) isotopes.
The oblate ground-state deformation found for Hg isotopes is
confirmed by experiment below $N=122$.
}
\end{figure}

The results for Sn isotopes are displayed in Fig.~\ref{fig:dvpn_50}
and those for Pb in Fig.~\ref{fig:dvpn_82}.
The dimensionless deformation parameter $\beta_2^{\text{calc}}$
is related to the intrinsic mass quadrupole moment of the
self-consistent mean-field wave functions
$\langle q | (2 z^2 - x^2 - y^2) | q \rangle$ as
\begin{equation}
\label{eq:beta:calc}
\beta_2^{\text{calc}}
\equiv \sqrt{ \frac{5}{16 \pi} } \, \frac{4 \pi}{3 R_0^2 A}
      \langle q | (2 z^2 - x^2 - y^2 ) | q \rangle
\end{equation}
with $R_0 \equiv 1.2 \, A^{1/3} \, \text{fm}$. In the three upper panels
of the figures, the results for isotopic chains with two protons less
are also displayed, such that the values for all four nuclei entering
$\delta V_{pn}$ are given in the same plot. The contribution of
deformation and correlation energy to
a given $\delta V_{pn}(N,Z)$ value can be extracted from the plot by first taking
the difference between the energies for $N$ and $N-2$ on the curves for
$Z$ and $Z-2$, and then subtracting the value for $Z-2$ from the one for
$Z$. To facilitate this comparison, the deformation and correlation
energies have been multiplied by a factor $1/4$, so that the quantity
entering $\delta V_{pn}$, Eq.~(\ref{eq:dVpn:def}) is plotted on the
figure. The magnitude of the contributions of deformation and correlations
to $\delta V_{pn}(Z,N)$ is directly related to the difference in slopes
of the curves for $Z$ or $Z-2$ for a given $N$. The largest contributions
are obtained when one of the slopes is much steeper than the other.

The ground-state configurations obtained for most Sn and Pb isotopes
are spherical; some mid-shell isotopes are slightly deformed.
However, the energy gain due to deformation in those cases is small,
smaller than 200~keV, and it originates from a deformed minimum nearly
degenerate with the spherical configuration. By contrast,
the deformation and gain in deformation energy for the ground states of
the non-magic
$Z-2$ isotopic chains can be large and vary rapidly for some
neutron numbers. For those cases, the contribution of the deformation
energy to $\delta V_{pn}$ is large.
This clearly indicates that one cannot assume to describe all the four
nuclei entering $\delta V_{pn}$ by a common mean-field, even for
closed-shell nuclei.

The correlation energy is larger for all Cd and Hg isotopes than for
Sn and Pb nuclei with same $N$.
However, the slopes of the $Z$ and $Z-2$ curves
differ significantly for a few isotopes only, and its contribution to
$\delta V_{pn}$ is large only for these neutron numbers. Nevertheless,
the beyond-mean-field correlations level out the rapidly fluctuating
effect of static
deformations. Their contribution to the binding energy plays also a
key role in the description of the two-proton separation energies across
the $Z=50$ and $Z=82$ shell~\cite{Ben06a,Ben08a}.

The agreement between the experimental data and the results of the
beyond-mean-field calculation is very satisfactory for the Sn and Pb
isotopic chains, as can be seen in the bottom panel of the
Figs.~\ref{fig:dvpn_50} and~\ref{fig:dvpn_82}. In particular,
only the latter calculation is able to describe the rise of $\delta V_{pn}$
up to $N=126$ and its sudden drop beyond.
The rapid variation of $\delta V_{pn}$ around \nuc{132}{Sn}
and \nuc{208}{Pb} is mainly due to the onset of substantial
beyond-mean-field correlations around doubly-magic nuclei.
This scenario is much more involved than the
proton-neutron interaction between the valence orbitals invoked
in Ref.~\cite{Bre10aM}. A detailed analysis of the contributions
to the $\delta V_{pn}$ value of \nuc{208}{Pb} is given in
Sect.~\ref{subsect:results:selectednuclei}.

\begin{figure}[t!]
\centerline{\includegraphics{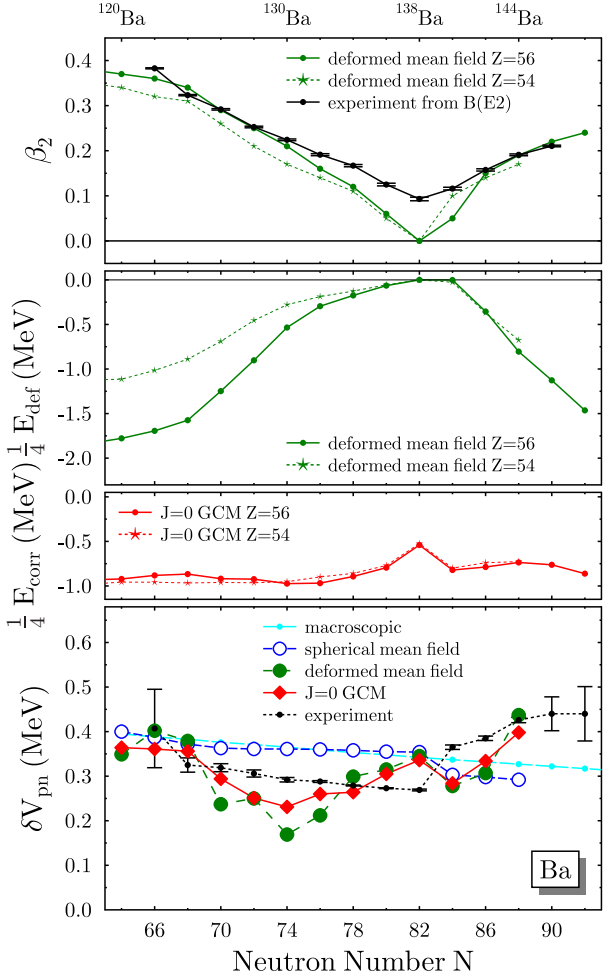}}
\caption{
\label{fig:dvpn__56}
(Color online)
Top to bottom: Intrinsic deformation of the mean-field
ground state compared with experimental data taken from \cite{nudat},
$1/4$ of the deformation energy, $1/4$ of the beyond-mean-field
correlation energy, and $\delta V_{pn}$ values obtained from the
macroscopic model, spherical
mean-field calculations, deformed mean-field calculations, $J=0$ projected
GCM calculations and experimental data for the chain of Ba ($Z=56$)
isotopes. For the deformation, deformation energy and correlation energy
the values for Xe ($Z=54$) isotopes are shown as well (see text).
}
\end{figure}

\begin{figure}[t!]
\centerline{\includegraphics{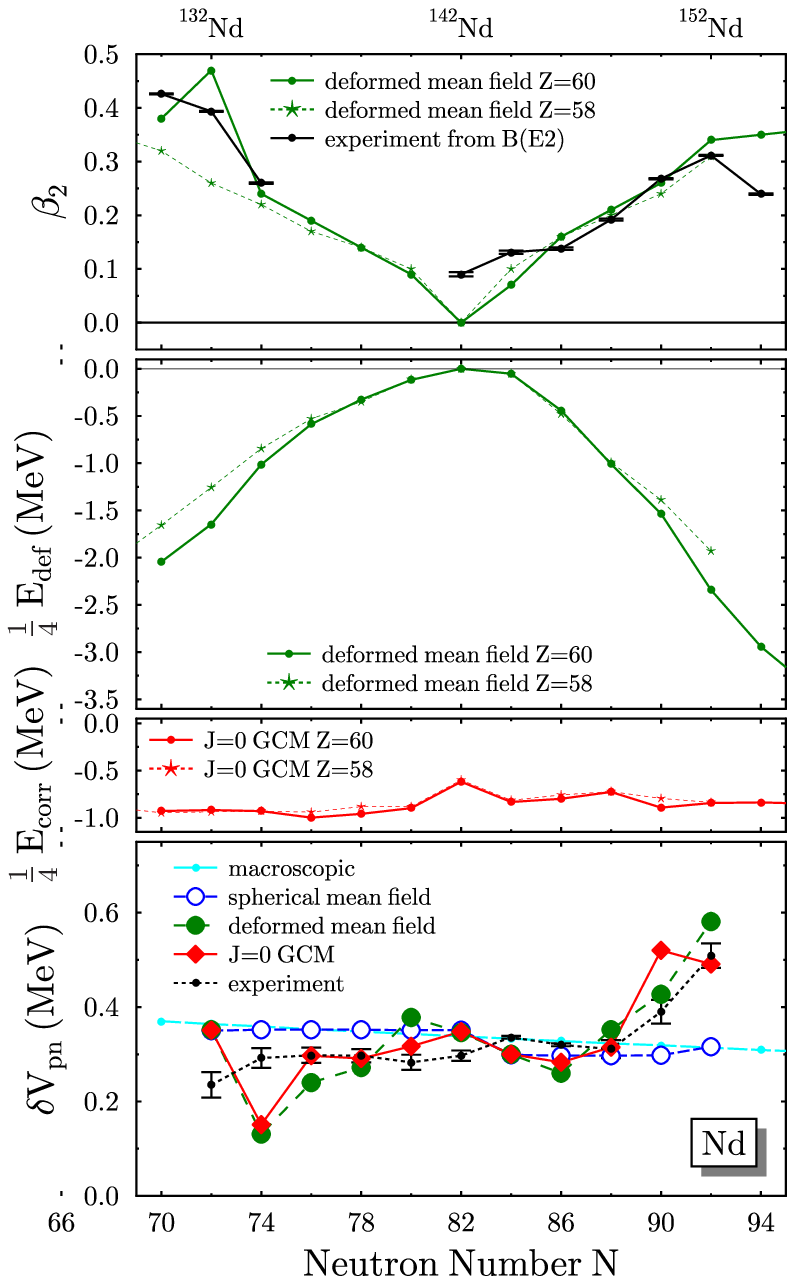}}
\caption{
\label{fig:dvpn__60}
(Color online)
The same as Fig.~\ref{fig:dvpn__56}, but for the chain of Nd
($Z=60$) isotopes.
}
\end{figure}

\begin{figure}[t!]
\centerline{\includegraphics{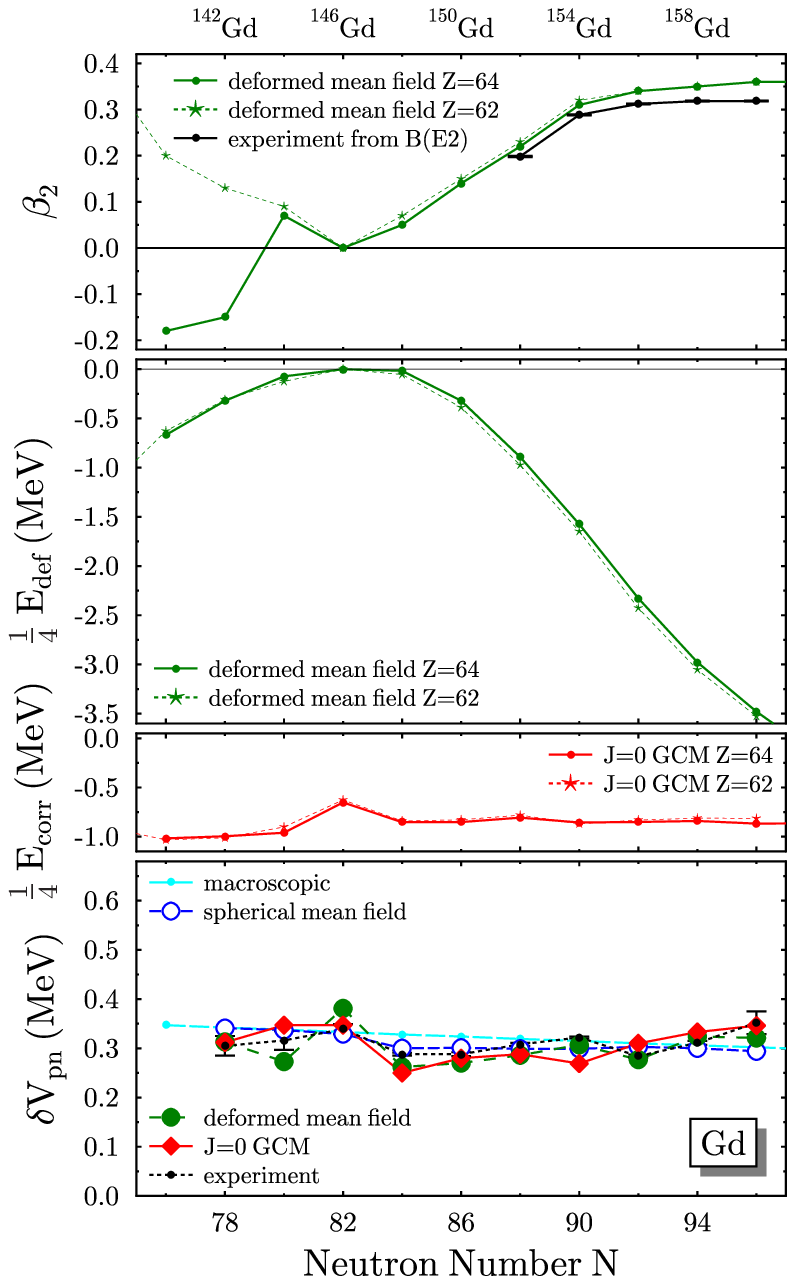}}
\caption{
\label{fig:dvpn__64}
(Color online)
The same as Fig.~\ref{fig:dvpn__56}, but for the chain of Gd
($Z=64$) isotopes.
}
\end{figure}

%
%
\subsubsection{Onset of deformation in rare-earth nuclei}

Let us now analyze the isotopic chains of Ba ($Z=56$), Nd ($Z=60$),
and Gd ($Z=64$), which cover a region of nuclei with a large variation
of deformation on both sides of the spherical $N=82$ shell closure.
Results of our calculations are compared with the experimental data in
Figs.~\ref{fig:dvpn__56},~\ref{fig:dvpn__60} and~\ref{fig:dvpn__64}.
Intrinsic deformations calculated with Eq.~(\ref{eq:beta:calc}) are
compared to values taken from Refs.~\cite{raman,nudat} that are
determined from experimental $B(E2)$ values making the assumption
of a rigid axial rotor
\begin{equation}
\label{eq:beta:expt}
\beta_2^{\text{expt}}
\equiv \frac{4 \pi}{3 Z R_0^2} \big[ B(E2, 0^+_1 \to 2^+_1) \big]^{1/2}
\end{equation}
with $R_0 \equiv 1.2 \, A^{1/3} \, \text{fm}$. For well-deformed nuclei,
theoretical~(\ref{eq:beta:calc}) and experimental~(\ref{eq:beta:expt})
values are in excellent agreement. Around spherical shell closures,
however, the lowest $2^+$ state is dominated either by non-collective
two-quasiparticle configurations or by fluctuations in collective degrees
of freedom, neither of which can be described by the mean-field
ground state.

For none of these three isotopic chains, the
spherical mean-field result for $\delta V_{pn}$ does show any
structure except for a tiny drop at the $N=82$ shell closure that
becomes rapidly smaller with increasing proton number. Besides that,
the spherical mean-field values remain
very close to the macroscopic ones for all three chains
of nuclei. The only isotopes to remain spherical when deformations are
allowed, are those with $N=82$. All heavier isotopes are prolate, with
a very similar variation of deformation as a function of $N$ for all
of the three isotopic chains.
Lighter isotopes are prolate for Ba and Nd, and oblate for Gd.
Although the deformation varies with $N$ in a rather similar way for the
three chains, the effects of deformation and beyond-mean-field correlations
on $\delta V_{pn}$ are different. Deformation and correlations do not
bring very large contributions, but they induce a significant change
of behavior of $\delta V_{pn}$ for Ba and Nd.
Note also that the changes
with respect to the spherical case bring theory closer to experiment with
a very few exceptions. The contribution from deformation energy
overcorrects the spherical result for $\delta V_{pn}$, in particular
by making $\delta V_{pn}$ smaller below $N=82$ and larger above this value. The
beyond-mean-field correlations straighten the curve and bring it very
close to the data.

Comparing the three isotopic chains, the largest deviation between
the experimental and the macroscopic $\delta V_{pn}$ values is observed
for the Ba ($Z=56$) isotopes. For the isotopes below $N=82$ (with the
possible exception of the lightest one $N=66$), the experimental values
are smaller, whereas above $N=82$ they are larger. The same overall
behavior is also found for Nd ($Z=60$), but with a smaller deviation
from the macroscopic results. For Gd ($Z=64$), the experimental data
lie almost on a straight line, very close to the macroscopic results.

The EDF models provide a simple explanation of these different
behaviors. The three  chains present a similar evolution as a function
of $N$, going from deformed to spherical to deformed shapes again.
However, looking to Fig.~\ref{fig:dvpn_def}, one can see that
the chains are located differently
with respect to the center of the deformed region.
The Ba ($Z=56$) isotopes are situated at the
lower end, where the deformation energy grows with $N$ at
a very different rate for $Z$ and $Z-2$ and brings a large
contribution to $\delta V_{pn}$. The Gd ($Z=64$) chain
is close to the center of the deformed region where the deformation
energy of adjacent isotones grows synchronously.
The $\delta V_{pn}$ values are not only unaffected by
deformation in the Gd isotopes, it is
also remarkable that the shape transition from a prolate shape
for $Z=62$ to an oblate one for $Z=64$ at $N=78$ does not
visibly affect the $\delta V_{pn}$ value obtained from the deformed mean-field
calculation. These examples indicate that $\delta V_{pn}$
cannot always be expected to be a sensitive indicator for changes in
deformation.

%
%
\subsubsection{$\delta V_{pn}$ along lines of constant $N+Z$ or $N-Z$}

\begin{figure}[t!]
\centerline{\includegraphics{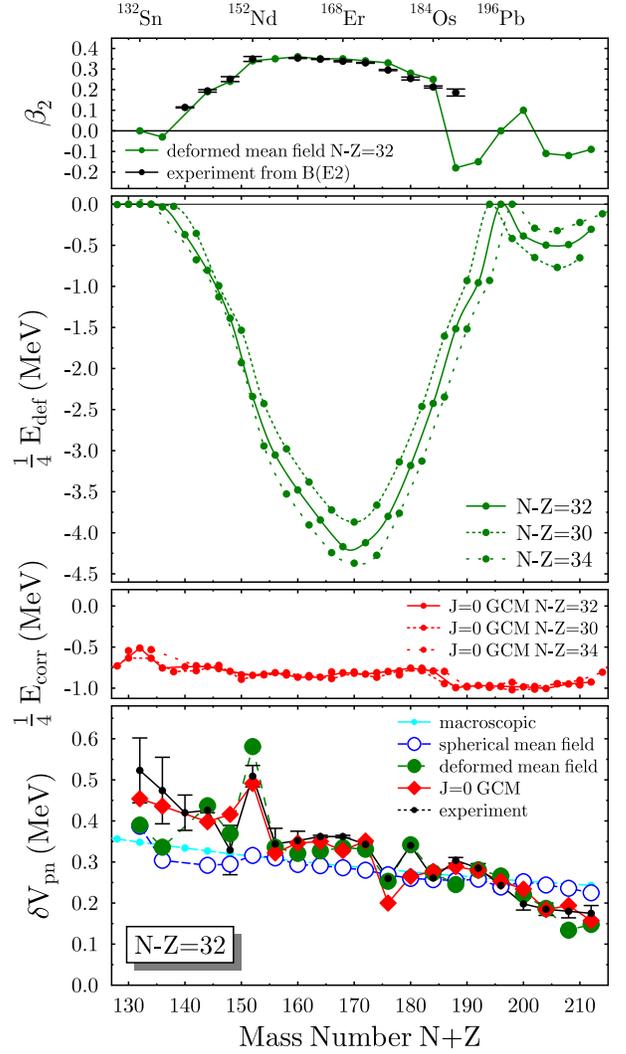}}
\caption{
\label{fig:dvpn_i32}
(Color online)
The same as Fig.~\ref{fig:dvpn__56}, but for the chain of constant
$N-Z=32$.
The nuclei for which data exist are expected to be prolate,
whereas the systematics of rotational bands and radii in the
Hg isotopic chain suggests that \nuc{194}{Hg} is oblate in its
ground state, in agreement with the calculation.
}
\end{figure}

\begin{figure}[t!]
\centerline{\includegraphics{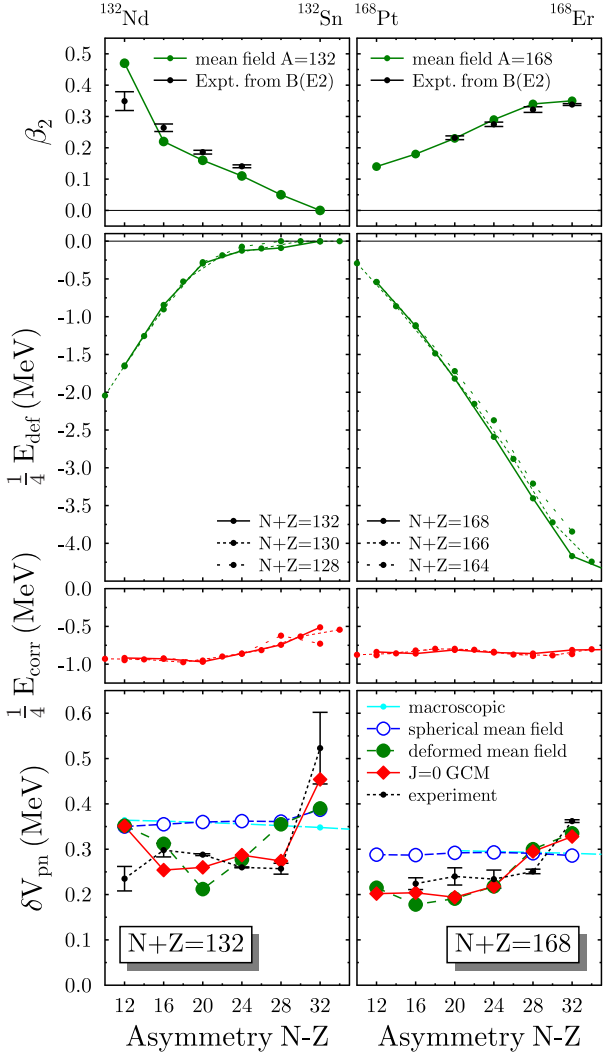}}
\caption{
\label{fig:dvpn_a132}
(Color online)
The same as Fig.~\ref{fig:dvpn__56}, but for the isobaric chains
with $A = N+Z = 132$ and $A = 168$.
}
\end{figure}

Let us now examine the evolution of $\delta V_{pn}$ along other cuts
through the chart of nuclei.
In Fig.~\ref{fig:dvpn_i32}, results for
nuclei on a line of constant $T_z = N-Z = 32$ are provided as a function
of $A = N+Z$. Available data start at the doubly-magic nucleus
\nuc{132}{Sn}, cover the entire rare-earth region, and extend beyond
$Z=82$. Most of the nuclei along this chain are deformed, except for
$Z=50$ and $82$. The deformation energy  takes its largest absolute
value of around 16.8~MeV for \nuc{168}{Er} at $Z=68$. However, it is
not the magnitude of the deformation energy that governs the size of
its contribution to $\delta V_{pn}$, but how the curve for $T_z = 32$
diverges from those for $T_z = 30$ and 34.
Indeed, the contribution
of the deformation energy to $\delta V_{pn}$ is obtained as the sum
of the energies of two successive points on the $T_z = 32$ line from which
one subtracts the sum of the values on the $T_z = 30$ and $T_z = 34$ lines corresponding to
the intermediate $A$ value.
The contributions of deformation and beyond-mean-field correlations
reinforce themselves for the lightest nuclei, leading to a rather
irregular pattern, significantly different from the structureless
spherical results. It is remarkable that $\delta V_{pn}$ values obtained
from the beyond-mean-field calculation follow very closely the many
irregularities of the experimental data.

Results obtained for two
nuclear chains corresponding to fixed values of $A$, which
are perpendicular to the $N-Z=32$ chain, are plotted
in Fig.~\ref{fig:dvpn_a132}. The first one corresponds to $A=132$
and extends from the very deformed neutron-deficient \nuc{132}{Nd}
to the doubly magic \nuc{132}{Sn}. As in the case of the Ba and Nd
isotopic chains, the different onset of deformation
and correlation energy for $Z$ and $Z-2$ lowers $\delta V_{pn}$
relative to the macroscopic values below the $N=82$ shell closure,
whereas it is enhanced for \nuc{132}{Sn}.
Note for this nucleus the significant differences between the
deformed and the beyond-mean-field calculations.
The change
of behavior for $N-Z = 32$ of $\delta V_{pn}$ with respect to the
spherical and macroscopic values is very nicely described by the
beyond-mean-field calculation.

The second isobaric chain in Fig.~\ref{fig:dvpn_a132},
$A=168$, almost follows the diagonal in
Fig.~\ref{fig:dvpn_def}. All isobars are deformed and the
deformation increases gradually when going from the very
neutron-deficient \nuc{168}{Pt} to \nuc{168}{Er}, a nucleus located
in the center of the deformed rare-earth
region. The deformed mean-field and beyond-mean-field calculations
give very similar $\delta V_{pn}$ values
and agree well with the data.
For the lighter isobars, the $\delta V_{pn}$ values are smaller
than the macroscopic ones, whereas for \nuc{168}{Er} they suddenly
increase to values above. For the chains that we discussed up to now,
the sudden increase of $\delta V_{pn}$ from below to above macroscopic
values took place when crossing a spherical shell closure,
i.e.\ with
decreasing deformation. In the case of the $A=168$ chain,
the sudden increase of $\delta V_{pn}$ has its origin in
the saturation of deformation energy with increasing
asymmetry.

Again, the deviation of $\delta V_{pn}$ from
the macroscopic value depends on the difference of increase in
deformation energy in adjacent nuclei.

%
%
\subsubsection{Doubly-magic nuclei and mutually enhanced magicity}

In Figs.~\ref{fig:dvpn_50},~\ref{fig:dvpn_i32}, and~\ref{fig:dvpn_a132},
the $\delta V_{pn}$ value of \nuc{132}{Sn} sticks out as being larger
than that of all surrounding nuclei. The same result is obtained for \nuc{208}{Pb}.
In both cases, these $\delta V_{pn}$ values are also much larger than the
macroscopic trend. A similar singular behavior for doubly-magic nuclei
is also found with other mass filters, such as two-particle separation
energies or $Q_\alpha$ values:  the value obtained for a doubly-magic
nucleus is much larger than those of
adjacent nuclei, including the semi-magic ones. This gives the impression
that the shell closure of one nucleon species reinforces the
magicity of the other, an effect sometimes called ``mutually
enhanced magicity'' in the literature~\cite{Schm79a,Zel83a,Lun03a}.
This effect is not described by pure mean-field models for which
two-nucleon separation energies or $Q_\alpha$ values across a shell
closure usually show very little variation with the number of
particles of the other species, in stark contrast to the data. In
Refs.~\cite{Fle04a,Ben06a,Ben08a}, it was shown that these filters are
much better described when beyond-mean-field
correlations are taken into account.
The same result is found here for
$\delta V_{pn}$. The beyond-mean-field correlation energy
is much smaller in a doubly-magic nucleus than in its neighbors.
Its rapid variation gives a contribution to $\delta V_{pn}$
that pushes it to very large values in doubly-magic nuclei, up to
twice as large as the average trend.

At the same time, the pattern of the $\delta V_{pn}$ values changes
for doubly-magic nuclei. For nuclei located either below or above the
shell closures for both nucleon species, the experimental
$\delta V_{pn}$ tends to be \emph{larger} than the average trend.
In contrast, for nuclei where one nucleon species is below and the
other above the respective shell closure, the experimental
$\delta V_{pn}$ value  tends to be \emph{smaller} than the
average trend. This behavior is very well illustrated in
Figs.~\ref{fig:dvpn__56},~\ref{fig:dvpn__60} and~\ref{fig:dvpn__64}
for the shell closures at $Z=50$ and $N=82$.

In the literature, qualitative
explanations have been proposed for this effect,
based on the nature of the orbitals filled by neutrons and protons.
If the energies of both orbitals are larger or smaller ($p-p$ and $h-h$ cases) than that of
the Fermi level,
they are supposed to
have a large overlap. On the contrary, if one of the energies is larger and the other smaller,
($p-h$ and $h-p$ case), this overlap is supposed to be small. The
behavior of $\delta V_{pn}$ is then attributed to the differences
between the overlaps~\cite{Cak06a}.
EDF calculations offer a different and more straightforward explanation.
This effect results from the combination
of a smooth macroscopic background and the contributions from
deformation and beyond-mean-field correlation energies. As can be seen
in ~Fig.~\ref{fig:dvpn_def}, their combined absolute value
increases in all directions around a doubly-magic nucleus. Moreover,
looking for instance at Figs.~\ref{fig:dvpn_50} and~\ref{fig:dvpn_82},
one sees that this increase is nonlinear.
The pattern that is observed for $\delta V_{pn}$ around doubly-magic
nuclei is then a trivial consequence of the asymmetry of the relative
signs of the four energies entering its definition.
For example,
for $N=118$, the $Z=74$ nucleus
is located in such a way that the
nonlinear increase of these contributions is pointing toward $N-2$,
$Z-2$. For the same value of $N$ but $Z=78$ the iso-energy
lines are nearly parallel to the $N$-axis and the contributions for a
given $Z$-value nearly cancel out.

Looking once more at Fig.~\ref{fig:dvpn_def}, the picture on how $\delta V_{pn}$
is build up emerges clearly. Let us divide the map into rectangles delimited by
the proton and neutron magic numbers.
In any of these rectangles, deformation and correlation energies grow nonlinearly
from small values along all borders to large ones in the middle.
The resulting map of deformation and correlation energies is highly
symmetric and centered around the middle of the region. The definition
of $\delta V_{pn}$, Eq.~(\ref{eq:dVpn:def}), however, is asymmetric.
It is designed to probe the increase of the energy when going from the
lower left to the upper right in the nuclear chart under the assumption
that it is superimposed on a background of like-particle interactions independent
on the number of the other particle. The combined
deformation and dynamical correlation energy does rarely follow this
anticipated pattern. As a consequence,
one always obtains positive contributions to $\delta V_{pn}$ around the
so-called $p-p$ and $h-h$ corners of the rectangle and
negative values in  the $p-h$ and $h-p$ corners.
Close to the center of such a major-shell
region, this trend is inverted when the deformation and correlation energies
reach their maximum. This explanation of the pattern of $\delta V_{pn}$
around shell closures does not invoke any knowledge about the spatial
structure of the single-particle orbits and their overlaps, and indicates
also that the observed pattern of $\delta V_{pn}$ does not necessarily
signal stronger or weaker proton-neutron interactions in the four corners
of a region of the nuclear chart between major shells.

The observation that the appearance of enhanced $\delta V_{pn}$
values in the rare-earth region when going from \nuc{132}{Sn} to
\nuc{208}{Pb} and beyond, is correlated to the line of
$N_{\text{valence}} \approx Z_{\text{valence}}$ has led the authors
of Ref.~\cite{Cak10a} to the speculation that this phenomenon might
be due to a ``mini-Wigner energy'' of origin similar to the Wigner
energy that leads to enhanced $\delta V_{pn}$ values along the
$N = Z$ nuclei. Our analysis makes this scenario very unlikely and
offers a simpler explanation.
First, we underline that our model does not give any trace of the
Wigner energy and its contribution to $\delta V_{pn}$
at the $N=Z$ line, as is the case for all present-day
self-consistent mean-field models~\cite{Sto07a,Sat97a,Per04a},
meaning that the relevant physics is not contained in it. In
contrast, our model does reproduce very well the enhanced
$\delta V_{pn}$ values along the
$N_{\text{valence}} \approx Z_{\text{valence}}$ line
in the rare-earth region. As explained above, their enhancement
is a consequence of the onset of deformation and beyond-mean-field
correlations when going away from a doubly-magic nucleus, which
gives a positive contribution to $\delta V_{pn}$ in some direction
and negative in other directions due to the asymmetric definition of
$\delta V_{pn}$.

%
%
\subsection{Detailed analysis of $\delta V_{pn}$ for selected nuclei}
\label{subsect:results:selectednuclei}

%
%
\subsubsection{General comments}

The discussion above demonstrates that the rapid variation of the
deformation energy and the beyond-mean-field correlation energy from symmetry
restoration and shape mixing often gives large contributions to
$\delta V_{pn} (N,Z)$. For nuclei away from the $N=Z$ line, this variation is at
the origin of almost all structures seen in the data. This also indicates
that the assumption of a common single-particle basis made in
Sect.~\ref{sect:theory:analysis} to obtain a simple expression for
$\delta V_{pn}$ in terms of proton-neutron matrix elements is rarely
justified. When the structure of the four nuclei entering
Eq.~(\ref{eq:dVpn:def}) is different, the question arises
whether
there are other terms in the energy functional than
the proton-neutron interaction energy that contribute to $\delta V_{pn}$.

We have selected three representative nuclei for which we will
decompose $\delta V_{pn} (N,Z)$ into contributions from the
proton-proton, neutron-neutron and proton-neutron terms in the EDF.

\begin{table}
\centering
\caption{
\label{dec}
Decomposition of $\delta V_{pn}$ into contributions coming from the
different terms of the energy density functional for \nuc{208}{Pb},
namely kinetic energy of neutrons and protons, the neutron-neutron,
proton-proton and proton-neutron parts of the Skyrme EDF, the
neutron-neutron and proton-proton parts of the pairing functional
and the proton-proton Coulomb EDF. We also give the sum of all
terms and the experimental value. All energies are in MeV.
}
\begin{tabular}{lcccc}
\hline \noalign{\smallskip}
                    & \multicolumn{4}{c}{$^{208}$Pb}  \\
\noalign{\smallskip} \cline{2-5} \noalign{\smallskip}
     term           & frozen HF & HF  & HF+BCS+LN & $J=0$ GCM \\
\noalign{\smallskip} \hline \noalign{\smallskip}
  kinetic n   &  0.000 &  -0.056 &  -0.010 &  -0.105 \\
  kinetic p   &  0.000 &   0.012 &  -0.044 &  -0.132 \\
  Skyrme  nn  &  0.025 &   0.076 &  -0.005 &   0.160 \\
  Skyrme  pp  &  0.017 &  -0.008 &   0.025 &   0.275 \\
  Skyrme  pn  &  0.162 &   0.211 &   0.218 &   0.382 \\
  pairing nn  &    -   &     -   &   0.027 &  -0.057 \\
  pairing pp  &    -   &     -   &  -0.005 &  -0.081 \\
  Coulomb     &  0.000 &   0.010 &   0.020 &   0.012 \\
\noalign{\smallskip} \hline \noalign{\smallskip}
  total       &  0.204 &   0.245 &   0.225 &   0.457 \\
\noalign{\smallskip} \hline \noalign{\smallskip}
$\delta V_{pn}^{\text{exp}}$  & -  & -  &         &   0.427 \\
\noalign{\smallskip} \hline
\end{tabular}
\end{table}

%
%
\subsubsection{\nuc{208}{Pb}}

The first nucleus \nuc{208}{Pb} has been chosen
for two reasons. First, we have seen in Fig.~\ref{fig:dvpn_82}
that the contribution of the beyond-mean-field correlation energy to
the $\delta V_{pn}$ value of this doubly-magic nucleus is particularly
large. Second, the spherical \nuc{208}{Pb} presents a very favorable
situation to numerically test the frozen HF approximation, where
the same set of single-particle wave functions is used to compute the
energy of all four nuclei involved in the computation of  $\delta V_{pn}$.
As discussed in Sect.~\ref{subsect:frozenHF},
this approximation has to be made to establish the direct relation
between $\delta V_{pn}$ and the two-body proton-neutron interaction.
In fact, \nuc{208}{Pb} is one of the very few spherical nuclei
for which such calculations can be performed. It requires that four
neighboring nuclei have a closed shell configuration, which is possible
only for $N$ and $Z$ values for which the orbitals below the Fermi
level are $p_{1/2^{-}}$ or $s_{1/2^{+}}$ levels for both protons and
neutrons. These conditions are met for \nuc{208}{Pb}, with a $\nu$ $s_{1/2^+}$
level below $N=126$ and a $\pi$ $p_{1/2^-}$ level below $Z=82$.

\begin{figure}[t!]
\centerline{\includegraphics{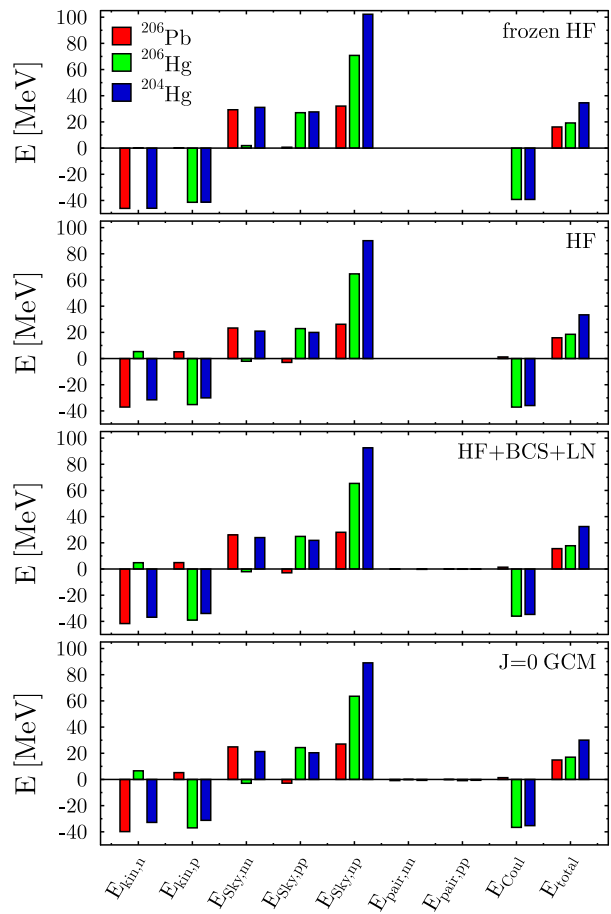}}
\caption{
\label{fig:pb208:decomp}
(Color online)
Difference between the values of the terms of the energy functional
listed in Table~\ref{dec} for \nuc{206}{Pb}, \nuc{206}{Hg} and \nuc{204}{Hg}
and their value for \nuc{208}{Pb}. The same four sets of calculations as
in Table~\ref{dec} are considered. All energies are in MeV.
}
\end{figure}

The results are presented in Table~\ref{dec}.
In the first column, the four nuclei entering $\delta V_{pn}$
have been calculated with the single-particle basis of $^{208}$Pb ("frozen HF"
approximation), without readjustment of the basis for each nucleus and
without pairing correlations. Self-consistency for \nuc{206}{Pb}, \nuc{206}{Hg}
and \nuc{204}{Hg} has been considered for the results given in
the second column ("HF"), and self consistency and
pairing correlations treated with the BCS+LN prescription have been taken
into account for the values of the third column.
Finally, the fourth column corresponds to the
$J=0$ projected GCM calculation.

We decompose the energy density functional into the kinetic energies of
neutrons and protons (including the centre-of-mass correction), the
neutron-neutron, proton-proton and proton-neutron parts of the
Skyrme EDF that models the particle-hole part of the effective strong
interaction, the neutron-neutron and proton-proton parts of the
pairing functional and the proton-proton Coulomb energy. The Skyrme
and pairing functionals contain density-dependent terms. We
interpret them as a density-dependence of the respective neutron-neutron,
proton-proton and proton-neutron terms. This choice of decomposition
is not unique, however. For further details about the functional, we
refer to Refs.~\cite{Ben03a,Ben02a}.

In the frozen HF calculation, the sole contribution to
$\delta V_{pn}$ comes from the Skyrme EDF. The neutron-proton
terms give the largest contribution, although the neutron-neutron
and proton-proton terms contribute by about $20 \%$ through their
density-dependence.
As soon as self-consistent wave functions are used, the one-body
contribution from the kinetic energy becomes large.
This is not surprising, as the kinetic energy
provides a large contribution to the symmetry energy coefficient
$a_{\text{sym}}$ of the EDF \cite{Ben02a,Sat06a}, which in turn
dominates the global trend of $\delta V_{pn}$, Eq.~(\ref{eq:mac:sym}).
All other terms in the functional are modified and can bring sizable
contributions to $\delta V_{pn}$. Pairing correlations change all
contributions even further.
The final value of $\delta V_{pn}$ from a self-consistent calculation
results from a partial cancelation of several terms.
The  proton-neutron terms in the Skyrme functional are of the right order
of magnitude and approach the final value of $\delta V_{pn}$
at a given level of approximation, but even for \nuc{208}{Pb},
which is probably one of the most favorable cases for the
frozen approximation, one can hardly conclude that $\delta V_{pn}$
is a valuable measure of proton-neutron interactions.

More importantly, the spherical mean-field values are far from the
experimental data and the correlations brought by symmetry restoration
and configuration mixing are large and crucial to obtain the correct
value, as they increase $\delta V_{pn}$ by 0.232 MeV to almost twice
the mean-field value. The decomposition of the $J=0$ projected GCM
results is listed in the fourth column of Table~\ref{dec}. All terms
become large in absolute value, but partially cancel each other.
Their sum gives a value for $\delta V_{pn}$ close to the data.
This is not entirely surprising as correlation energy is
always gained from the compensation between a loss in kinetic energy
and a gain in interaction energy. Table~\ref{dec} indicates that
there are large contributions to $\delta V_{pn}$ from proton-proton,
neutron-neutron and neutron-proton terms and that none of them is
dominant.

Fig.~\ref{fig:pb208:decomp} illustrates how the  $\delta V_{pn}$
is build up from cancelations between the contributions of the four nuclei
entering its definition. It shows the differences between the
EDF terms defined in Table~\ref{dec} for \nuc{206}{Pb},
\nuc{206}{Hg} and \nuc{204}{Hg} and their value in \nuc{208}{Pb}.
The contribution of a given term to $\delta V_{pn}$(\nuc{208}{Pb})
is the sum of the values for \nuc{206}{Pb} (red) and \nuc{206}{Hg}
(green) minus the value for \nuc{204}{Hg} (blue), divided by four.
While in the frozen HF approximation, most terms cancel out,
 exactly self-consistency, pairing and collective quadrupole correlations
lead to a much more complex situation. One can conclude from this
analysis that each contribution to $\delta V_{pn}$ itself is the
result of a partial compensation between changes in all four nuclei
entering its definition.

%
%
\subsubsection{Deformed rare-earth nuclei}

\begin{table}
\centering
\caption{
\label{dec2}
Decomposition of $\delta V_{pn}$ values for \nuc{168}{Er} and
\nuc{152}{Nd} into contributions coming from the
different terms in the energy density functional in deformed
self-consistent mean-field calculations,
namely kinetic energy, the neutron-neutron,
proton-proton and proton-neutron parts of the Skyrme EDF,
pairing energy and the proton-proton Coulomb EDF.
We also give the total contribution from beyond-mean-field
correlations, the sum of all these terms and the experimental
value. All energies are in MeV.
}
\begin{tabular}{lrr}
\hline \noalign{\smallskip}
    term    & $^{168}$Er & $^{152}$Nd \\
\noalign{\smallskip} \hline \noalign{\smallskip}
    kinetic            & -0.252      & -0.334 \\
    Skyrme nn          &  0.164      &  0.304 \\
    Skyrme pp          &  0.065      & -0.016 \\
    Skyrme np          &  0.372      &  0.676 \\
    Coulomb            &  0.005      & -0.161 \\
    pairing            & -0.020      &  0.081 \\
\noalign{\smallskip} \hline \noalign{\smallskip}
    total mean field  &  0.334 & \rule{0.25cm}{0.0cm} 0.550 \\
\noalign{\smallskip} \hline \noalign{\smallskip}
    correlation energy  \rule{0.35cm}{0.0cm} & -0.006     & -0.051 \\
\noalign{\smallskip} \hline \noalign{\smallskip}
$\delta V_{pn}^{theo}$ &  0.328     & 0.490 \\
\noalign{\smallskip} \hline \noalign{\smallskip}
$\delta V_{pn}^{exp}$  &  0.362     & 0.509 \\
 \noalign{\smallskip} \hline
  \end{tabular}
\end{table}

Let us now consider two nuclei in
a deformed region of the nuclear chart. The
four nuclei entering the calculation of $\delta V_{pn}$ for
\nuc{168}{Er} have similar deformations. By contrast, they differ
significantly for \nuc{152}{Nd}, as can be seen in Fig.~\ref{fig:dvpn__60}.
In both cases, the spherical mean-field results
are of no interest and will not be discussed here.
The contribution of correlations brought by configuration
mixing and symmetry restoration is also small and
the analysis of $\delta V_{pn}$ can be performed for
the deformed calculations only.
The value of $\delta V_{pn}$ results
from contributions coming from all terms in the functional.
The kinetic energy brings a large negative contribution, even
for \nuc{168}{Er} for which the four nuclei have very similar
deformations. This demonstrates that the frozen approximation is
not valid and that the use of a unique single-particle basis is
not justified. The largest positive contribution to $\delta V_{pn}$
comes from the Skyrme EDF. Other terms are small for \nuc{168}{Er},
but there is a large negative contribution of the Coulomb term in \nuc{152}{Nd}.
Note also that, although the contribution of correlations
is small for this nucleus, it has the right sign to bring the theoretical $\delta V_{pn}$
in good agreement with the data.

Altogether, these examples indicate
that, in a realistic model, $\delta V_{pn}$ is not determined
by the interaction between the last two valence nucleons, but has
contributions from the modifications of all single-particle
wave functions on the one hand, and from all terms in the
energy functional on the other.

%
%
\section{Discussion and Conclusions}
\label{sect:summary}

In this paper, we have analyzed in details the relevance of a
difference between the binding energy of four nuclei, called
$\delta V_{pn}$, as a measure of the proton-neutron interaction between
valence particles and as an indicator for structural changes
in nuclei.

We have first investigated whether one can derive a relation between
$\delta V_{pn}$ and a proton-neutron matrix element in simple models,
where analytic formula can be derived. Even in the oversimplified
case where the four nuclei entering $\delta V_{pn}$ can be described
by  HF wave functions generated by the same mean field and a two-body
interaction, one obtains only a relation between  $\delta V_{pn}$
and a combination of two matrix elements. Any higher-order term
in the interaction, such as a density-dependence or a three-body interaction,
complicates the relation, as do self-consistency and any correlation
such as pairing, deformation, or any configuration mixing.
This formal analysis already indicates that it can hardly
relate $\delta V_{pn}$ to any specific proton-neutron interaction
in a realistic model. This formal analysis is
confirmed by the decomposition of calculated $\delta V_{pn}$ values
for the doubly-magic spherical \nuc{208}{Pb} nucleus
and the deformed \nuc{152}{Nd} and \nuc{168}{Er}, for which we find
indeed that all terms in the energy functional contribute, not just the
proton-neutron interaction, and that self-consistency and correlations
beyond the mean field play a substantial role.

We have then shown that our beyond-mean-field method has all the
necessary ingredients to reproduce the global trends of $\delta V_{pn}$.
As has been pointed out earlier~\cite{Sto07a}, the global trend of
$\delta V_{pn}$ is determined by the symmetry and surface symmetry
energy coefficients that can be deduced from an energy functional.
With a detailed analysis of a few representative regions in
the nuclear chart, we have illustrated how the fine structure of
$\delta V_{pn}$ builds up from the successive introduction of
deformation and correlations due to symmetry restorations and
configuration mixing. Both are crucial ingredients for
the reproduction of data. As found earlier for two-nucleon separation
energies \cite{Ben05a,Ben06a,Ben08a}, within our model it is essential
to take into account beyond-mean-field correlations for the description of
data in the vicinity of magic numbers. We have checked that the large-scale
calculation of even-even nuclei using a mapped five-dimensional
microscopic Bohr-Hamiltonian based on the Gogny-force~\cite{Del10a}
gives qualitatively the same results as ours.

Our model provides a satisfactory description of the data, except at
the $N=Z$ line where the Wigner effect is absent from our results. Within the
framework of our model, however, $\delta V_{pn}$ does neither
provide a reliable \emph{measure} of the proton-neutron interaction
terms in the energy functional used, nor a reliable \emph{indicator}
for structural changes. Certainly, in some instances structural changes
such as the onset of deformation lead to anomalies in the $\delta V_{pn}$ values,
but in many other instances they do not, and there is no one-to-one
correspondence between a structural change and the resulting modification
of $\delta V_{pn}$. In turn, a local increase or decrease of
$\delta V_{pn}$ can have many different origins. One of the main limiting
factors for the use of $\delta V_{pn}$ as an indicator for structural
change is its asymmetric definition, which results in the fact that additional binding
from a change in nuclear structure contributes differently to
$\delta V_{pn}$ in sign and size depending on the direction in the
nuclear chart in which the structure changes.
In particular the characteristic pattern of $\delta V_{pn}$ having
a significantly larger size for nuclei where only particles or
only holes are added to a doubly-magic nucleus as compared to systems with particles added
for one nucleon species and holes for the other (a pattern recently
interpreted as a ``mini-Wigner energy'' \cite{Cak10a}),
is a trivial consequence of the asymmetric definition of $\delta V_{pn}$,
and \emph{not} an indicator for a qualitative difference in either the
proton-neutron interaction or a difference of their
structure.

The usefulness of $\delta V_{pn}$ is compromised by being a mass filter
of very high order that is thereby prone to unpredictable
cancelations when the nuclei entering $\delta V_{pn}$ have different
structure. Lower-order mass filters such as two-nucleon
separation energies or $Q$ values are usually more reliable
indicators of structural changes than $\delta V_{pn}$, although these
may fail as well. This also means that $\delta V_{pn}$ does not
provide a conclusive benchmark for nuclear EDF methods that would be superior
or complementary to other mass filters such as two-nucleon separation
energies and $Q$ values.

\begin{acknowledgments}

This work was supported by the ``Interuniversity Attraction Pole''
(IUAP) of the Belgian Scientific Policy Office under project P6/23.
The computations were performed at the National Energy Research Scientific
Computing Center, supported by the U.S. Department of Energy under
contract no. DE-AC03-76SF00098 and using HPC resources from GENCI-IDRIS
(Grant 2010-050707).
The authors also thank R.V.F. Janssens and
V.~Hellemans  for their critical reading of the manuscript.

\end{acknowledgments}

%
%

\end{document}